\documentclass[12pt]{article}

\usepackage{amsmath,mathtools,amsfonts}
\usepackage{amssymb}
\usepackage{bm}
\usepackage[dvips]{graphicx}
\usepackage{cite}

\newcommand{\sectiono}[1]{\section{#1}\setcounter{equation}{0}}

\newcommand{\llangle}{\langle\!\langle}
\newcommand{\rrangle}{\rangle\!\rangle}
\newcommand{\bd}[1]{\boldsymbol{#1}}
\newcommand{\id}{\mathbb{I}}

\newcommand{\idSH}{\mathbb{I}_{\mathcal{SH}}}
\newcommand{\NSNS}{NS\textrm{-}NS}
\newcommand{\RNS}{R\textrm{-}NS}
\newcommand{\NSR}{NS\textrm{-}R}
\newcommand{\RR}{R\textrm{-}R}

\begin{document}

\baselineskip=17pt

\begin{titlepage}
\rightline{\tt YITP-19-102}
\rightline\today
\begin{center}
\vskip 2.5cm
{\Large \bf {Type II superstring field theory
 with cyclic $L_\infty$ structure}}
\vskip 1.0cm
{\large {Hiroshi Kunitomo${}^1$ and Tatsuya Sugimoto${}^2$}}
\vskip 1.0cm
{\it {Center for Gravitational Physics}}, 
{\it {Yukawa Institute for Theoretical Physics}}\\
{\it {Kyoto University}},
{\it {Kyoto 606-8502, Japan}}\\
${}^1$kunitomo@yukawa.kyoto-u.ac.jp,\ ${}^2$tatsuya.sugimoto@yukawa.kyoto-u.ac.jp

\vskip 2.0cm

{\bf Abstract}
\end{center}

\noindent
We construct a complete type II superstring field theory that includes all
the NS-NS, R-NS, NS-R and R-R sectors. 
As in the open and heterotic superstring cases, 
the R-NS, NS-R and R-R string fields are constrained by using the 
picture-changing operators. In particular, we use a non-local inverse 
picture-changing operator for the constraint on the R-R string field, which seems
to be inevitable due to the compatibility of the extra constraint 
with the closed string constraints.
The natural symplectic form in the restricted Hilbert space gives
a non-local kinetic action for the R-R sector, but it correctly provides
the propagator expected from the first-quantized formulation.
Extending the prescription previously obtained for the heterotic string field theory,
we give a construction of general type II superstring products, 
which realizes a cyclic $L_\infty$ structure, and thus provides a 
gauge-invariant action based on the homotopy algebraic formulation. 
Three typical four-string amplitudes derived from the constructed 
string field theory are demonstrated to agree with those in 
the first-quantized formulation.
We also give the half-Wess-Zumino-Witten action defined in the medium
Hilbert space whose left-moving sector is still restricted to the small 
Hilbert space.

\end{titlepage}

\tableofcontents

\newpage

\sectiono{Introduction}\label{Introduction}

In recent years there has been some significant progress in 
constructing gauge-invariant superstring field theories. 
First, a complete WZW-like action  for open superstring including both 
the Neveu-Schwarz (NS) and Ramond (R) sectors, representing space-time bosons 
and fermions, respectively, has been constructed \cite{Kunitomo:2015usa}
after several significant developments 
\cite{Erler:2013xta,Erler:2014eba,Erler:2015rra,Erler:2015lya,Berkovits:1995ab,
Berkovits:2001im,Berkovits:2004xh,Kunitomo:2013mqa,Kunitomo:2014hba,Jurco:2013qra,
Matsunaga:2014wpa,Goto:2015pqv}.
A crucial idea to successfully incorporate the Ramond sector is
to impose an extra constraint on the Ramond string field, 
which can naturally be interpreted to come from the fermionic moduli 
integration over the super-Riemann surface.
Shortly thereafter, this was extended to an alternative formulation
in the small Hilbert space \cite{Erler:2016ybs}, 
in which the gauge symmetry is beautifully 
realized using a homotopy algebraic structure, the $A_\infty$ algebra.
Several interesting studies, such as on the general structure of
the complete WZW-formulation
\cite{Matsunaga:2015kra,Erler:2017onq}, on the space-time supersymmetry 
\cite{Erler:2016rxg,Kunitomo:2016kwh}, 
and on some generalization toward a heterotic string field theory
\cite{Goto:2016ckh,Kunitomo-HRI,Kunitomo:2019did},
have also been undertaken.
Then, in a previous paper, the authors extended these constructions
to the heterotic string field theory \cite{Kunitomo:2019glq}.
We first constructed a gauge-invariant action in the small
Hilbert space by constructing string interactions realizing a homotopy
algebraic structure of closed string, cyclic $L_\infty$ algebra,
and then also gave the WZW-like action through a field redefinition.

Independent of these developments, 
Ashoke Sen has developed a closed superstring 
field theory applicable to both the heterotic and type II theories 
\cite{Sen:2015hha,Sen:2015uaa}.
He has provided a quantum master action in a rather abstract way by considering 
string off-shell amplitudes allowing a cell decomposition.
In addition, instead of imposing constraint as in the former two formulations, 
he introduced an extra string field, which becomes free 
and decouples from the physical sector, to incorporate the Ramond sector
consistently. It has also been shown that it can also be extended to
the open superstring field theory \cite{Konopka:2016grr}.

Thanks to these developments, we now have three independent formulations 
of superstring field theory: the homotopy algebraic, the WZW-like, 
and the Sen's formulations. Each of these formulations has advantages
and disadvantages, and they seem to be complementary.
So the aim of this paper is to fill the blank still remaining by
constructing a complete field theory of the type II superstring
based on the homotopy algebraic and WZW-like formulations to provide
a solid foundation for non-perturbative studies of the superstring theories.

This paper is organized as follows. In section \ref{homotopy} we 
summarize how the type II superstring field theory is constructed based 
on the homotopy algebraic structure for the closed string, 
the cyclic $L_\infty$ algebra.
We impose constraints on the string fields in the R-NS, NS-R and R-R sectors. 
In the R-R sector, in particular,
we introduce non-local inverse picture-changing operator, 
which seems to be inevitable
due to the compatibility of the extra constraint 
with the closed string constraints. 
We construct the free theory and explain that it provides 
the correct R-R propagator even though the kinetic term is non-local.
We show that it can be replaced with the local action if an extra string
field is introduced following Sen's formulation. 
Then, it is shown that we can construct a gauge-invariant action 
if the string products have the cyclic $L_\infty$ structure.
Such string products are explicitly constructed in section
\ref{cyclic products}.
The prescription is an extension of the asymmetric construction proposed in 
Ref.~\cite{Erler:2014eba} for the NS-NS sector, and is obtained by repeating twice
the one proposed for the heterotic string products
in Ref.~\cite{Kunitomo:2019glq}.
The operators with non-zero picture number are inserted first 
for the left-moving sector and then for the right-moving sector,
following the procedure used for the heterotic string field theory.
We confirm that the action constructed in this way actually 
reproduces typical four-point amplitudes in section \ref{Four point}.
We explicitly calculate three on-shell four-point amplitudes with
four R-R; two NS-R, two R-R; and NS-NS, R-NS, NS-R, R-R external states, and show
that they agree with those obtained in the first-quantized formulation.
In section \ref{relation to WZW}, we attempt to map the action
and gauge transformation to those based on the WZW-like formulation.
Unfortunately, however, we only obtain a half-WZW-like action defined 
not in the large Hilbert space but in the medium Hilbert space,  
the tensor product of the large Hilbert space for the left-moving 
sector and the small Hilbert space for the right-moving sector.
Section \ref{summary} is devoted to the summary and discussion.
We summarize how the string field is expanded with respect 
to the ghost zero modes for each sector in appendix \ref{ghost zero modes},
which is useful in considering the Batalin-Vilkovisky 
(BV) quantization \cite{Batalin:1984jr}.

\sectiono{Type II string field theory in homotopy algebraic formulation}
\label{homotopy}

We first summarize several basics of type II string field theory
in the homotopy algebraic formulation. The free field theory is given and
discussed in some detail. After confirming that the action of the R-R sector
provides the propagator used in the first-quantized formulation,
we show that it can also be written in the local form by introducing an extra
R-R string field following Sen's formulation. 
The gauge-invariant interacting action 
can be obtained if we assume multi-string products with the cyclic 
$L_\infty$ structure.

\subsection{String field and constraints}\label{field and constraint}

There are four sectors, the NS-NS, R-NS, NS-R and R-R sectors,
in the first-quantized Hilbert space of the type II superstring, $\mathcal{H}$\,, 
\begin{equation}
 \mathcal{H}\ =\ \mathcal{H}_{\NSNS} 
+ \mathcal{H}_{\RNS} + \mathcal{H}_{\NSR} 
+ \mathcal{H}_{\RR}\,,
\end{equation}
corresponding to the combinations 
of the Neveu-Schwarz and Ramond boundary conditions for the left- 
and right-moving fermionic coordinates:
\begin{alignat}{3}
 \mathcal{H}_{\NSNS}\ =&\ \mathcal{H}^{NS}\otimes\bar{\mathcal{H}}^{NS}\,,
\qquad&
 \mathcal{H}_{\RNS}\ =&\ \mathcal{H}^{R}\otimes\bar{\mathcal{H}}^{NS}\,,
\nonumber\\
 \mathcal{H}_{\NSR}\ =&\ \mathcal{H}^{NS}\otimes\bar{\mathcal{H}}^{R}\,,
\qquad&
 \mathcal{H}_{\RR}\ =&\ \mathcal{H}^{R}\otimes\bar{\mathcal{H}}^{R}\,,
\end{alignat}
where $\mathcal{H}$ and $\bar{\mathcal{H}}$ on the right-hand sides are
the left-moving and right-moving small Hilbert spaces, respectively.
Accordingly, the type II string field $\Phi$ has four components,
\begin{equation}
 \Phi\ =\ \Phi_{\NSNS} + \Phi_{\RNS} 
+ \Phi_{\NSR} + \Phi_{\RR}
\in \mathcal{H}\,,
\end{equation}
which is Grassmann even and has ghost number $2$\,.
The first and the last components, $\Phi_{\NSNS}$ and 
$\Phi_{\RR}$\,, have picture number $(-1,-1)$ and $(-1/2,-1/2)$\,,
respectively, and represent space-time bosons.
The second and the third components, $\Phi_{\RNS}$ and
$\Phi_{\NSR}$\,, have picture number $(-1/2,-1)$ and $(-1,-1/2)$\,,
respectively, and represent space-time fermions.
All of these components satisfy the closed string constraints,
\begin{equation}
 b_0^-\Phi\ =\ L_0^-\Phi\ =\ 0\,,
\label{restrict closed}
\end{equation}
where $b_0^\pm=b_0\pm\bar{b}_0$\,, $L_0^\pm=L_0\pm\bar{L}_0$
and $c_0^{\pm}=(c_0\pm\bar{c}_0)/2$\,.
The first constraint imposes that the string field does not depend
on the ghost-zero mode $c_0^-$\,.
Therefore, the NS-NS component, in which
only the $bc$ ghosts have zero modes, is expanded with respect to
the ghost zero mode as
\begin{equation}
 \Phi_{\NSNS}\ =\ \phi_{\NSNS} - c_0^+\psi_{\NSNS}\,.
\label{restrict NSNS}
\end{equation}

As in the open and heterotic superstring field theories
 \cite{Kunitomo:2015usa,Erler:2016ybs,Kunitomo:2019glq},
we further restrict the dependence of the other components
on the $\beta\gamma$ ghost zero modes.
For the $\Phi_{\RNS}$ and $\Phi_{\NSR}$ components,
we impose
\begin{equation}
  XY\Phi_{\RNS}\ =\ \Phi_{\RNS}\,,\qquad
 \bar{X}\bar{Y}\Phi_{\NSR}\ =\ \Phi_{\NSR}\,,
\label{restrict RNS}
\end{equation}
respectively, where $XY$ and $\bar{X}\bar{Y}$ are the projection operators
defined by using the picture-changing operators and their inverses,
\begin{subequations} \label{PCO}%
 \begin{alignat}{3}
X\ =&\ -\delta(\beta_0)G_0 + \delta'(\beta_0)b_0\,,\qquad& 
Y\ =&\ -2c_0^+\delta'(\gamma_0)\,,\\
\bar{X}\ =&\ -\delta(\bar{\beta}_0)\bar{G}_0 + \delta'(\bar{\beta}_0)\bar{b}_0\,,
\qquad& \bar{Y}\ =&\ -2c_0^+\delta'(\bar{\gamma}_0)\,,
\end{alignat}
\end{subequations}
which satisfy the relations
\begin{subequations} \label{relation XYX}%
\begin{alignat}{5}
 XYX\ =&\ X\,,\qquad& YXY\ =&\ Y\,,\qquad & [\,Q\,, X\,]\ =&\ 0\,,
\label{relations holo}\\ 
 \bar{X}\bar{Y}\bar{X}\ =&\ \bar{X}\,,\qquad& 
\bar{Y}\bar{X}\bar{Y}\ =&\ \bar{Y}\,,\qquad& [\,Q\,, \bar{X}\,]\ =&\ 0\,.
\label{relations anti-holo}
\end{alignat}
\end{subequations}
Here $G_0$  and $\bar{G}_0$ are the zero modes of the left- and right-moving 
(total) superconformal currents, 
respectively.
Note that the inverse picture-changing operators in Eq.~(\ref{PCO}) are defined
so that the additional constraints in Eq.~(\ref{restrict RNS}) are compatible
with the closed string constraints of Eq.~(\ref{restrict closed}).
Since the picture-changing operators $X$ and $\bar{X}$ are BRST invariant,
they can be written as the BRST exact form in the large Hilbert space:
\begin{equation}
 X\ =\ [Q\,,\,\Xi]\,,\qquad \bar{X}\ =\ [Q\,,\,\bar{\Xi}]\,,
\end{equation}
with\footnote{In this paper, for notational simplicity,
we denote the zero modes of $\eta(z)$ and $\xi(z)$ as $\eta$ and $\xi$\,,
and those of $\bar{\eta}(\bar{z})$ and $\bar{\xi}(\bar{z})$ as $\bar{\eta}$ and
$\bar{\xi}$\,, respectively.}
\begin{subequations} 
\begin{align}
\Xi\ =&\ \xi+(\Theta(\beta_0)\eta\xi-\xi)P_{-3/2}
+(\xi\eta\Theta(\beta_0)-\xi)P_{-1/2}\,,\\
 \bar{\Xi}\ =&\ \bar{\xi}+(\Theta(\bar{\beta}_0)\bar{\eta}\bar{\xi}-\bar{\xi})\bar{P}_{-3/2}
+(\bar{\xi}\bar{\eta}\Theta(\bar{\beta}_0)-\bar{\xi})\bar{P}_{-1/2}\,.
\end{align}
\end{subequations}
The ghost zero-mode dependence of the components $\Phi_{\RNS}$ and $\Phi_{\NSR}$ 
is restricted to the form
\begin{subequations} \label{restricted RNS and NSR}
\begin{align}
\Phi_{\RNS}\ =&\ \phi_{\RNS} - \frac{1}{2}(\gamma_0 + 2c_0^+G)\psi_{\RNS}\,,\\
\Phi_{\NSR}\ =&\ \phi_{\NSR} - \frac{1}{2}(\bar{\gamma}_0+2c_0^+\bar{G})\psi_{\NSR}\,,
\end{align}
\end{subequations}
where $G=G_0+2\gamma_0b_0$ and $\bar{G}=\bar{G}_0+2\bar{\gamma}_0\bar{b}_0$ are  
the ghost zero-mode independent part of $G_0$ and $\bar{G}_0$\,, respectively.

On the other hand, for the $\Phi_{\RR}$ component which depends on
both the left- and right-moving $\beta\gamma$ zero modes,
we cannot simultaneously impose two conditions,
\begin{equation}
XY\Phi_{\RR}=\Phi_{\RR}\,,\qquad\bar{X}\bar{Y}\Phi_{\RR}=\Phi_{\RR}\,,
\nonumber
\end{equation} 
due to their non-commutativity: $[XY,\bar{X}\bar{Y}]\ne0$\,. 
However, we should notice that 
the choices of inverse picture-changing operators are not unique.
There is a possibility to use the non-local operators,
\begin{equation}
 \mathcal{Y}\ =\ -2\frac{G}{L_0^+}\delta(\gamma_0)\,,\qquad
 \bar{\mathcal{Y}}\ =\ -2\frac{\bar{G}}{L_0^+}\delta(\bar{\gamma}_0)\,,
\label{alternative Y}
\end{equation}
which also satisfy
\begin{subequations} \label{relation cal XYX} 
 \begin{align}
 X\mathcal{Y}X\ =&\ X\,,\qquad \mathcal{Y}X\mathcal{Y}\ =\ \mathcal{Y}\,,\\
 \bar{X}\bar{\mathcal{Y}}\bar{X}\ =&\ \bar{X}\,,\qquad 
\bar{\mathcal{Y}}\bar{X}\bar{\mathcal{Y}}\ =\ \bar{\mathcal{Y}}\,,
\end{align}
\end{subequations}
as the inverse picture-changing operators \cite{Jurco:2013qra}. 
We can impose the conditions
\begin{equation}
  X\mathcal{Y}\Phi_{RR}\ =\ \bar{X}\bar{\mathcal{Y}}\Phi_{RR}\ =\ \Phi_{RR}\,,
\label{constraint RR}
\end{equation}
which are now compatible
with each other, and also with the closed string constraints of 
Eq.~(\ref{restrict closed}).
It can be shown that 
the ghost zero-mode dependence of $\Phi_{\RR}$ 
is restricted by the constraints in Eq.~(\ref{constraint RR}) as
\begin{equation}
\Phi_{\RR}\ =\ \phi_{\RR} -\frac{1}{2}(\gamma_0\bar{G}
-\bar{\gamma}_0G+2c_0^+G\bar{G})\psi_{\RR}\,.
\label{restricted RR}
\end{equation}
Here we define $\psi_{\RR}$ so that the expansion in Eq.~(\ref{restricted RR}) has
a local form, which will be found to be natural shortly.

Here, if we define the operators $\mathcal{G}$ and $\mathcal{G}^{-1}$ by
\begin{subequations} 
 \begin{align}
 \mathcal{G}\ =&\ \id\pi_1^{(0,0)} + X\pi_1^{(1,0)} 
+ \bar{X}\pi_1^{(0,1)} + X\bar{X}\pi_1^{(1,1)}\,\\
 \mathcal{G}^{-1}\ =&\ \id\pi_1^{(0,0)} + Y\pi_1^{(1,0)} 
+ \bar{Y}\pi_1^{(0,1)} + \mathcal{Y}\bar{\mathcal{Y}}\pi_1^{(1,1)}\,,
\end{align}
\end{subequations}
the relations in Eqs.~(\ref{relation XYX}) and (\ref{relation cal XYX}) can collectively
be written as
\begin{equation}
 \mathcal{G}\mathcal{G}^{-1}\mathcal{G}\ =\ \mathcal{G}\,,\qquad
 \mathcal{G}^{-1}\mathcal{G}\mathcal{G}^{-1}\ =\ \mathcal{G}^{-1}\,,\qquad
 [\,Q\,,\mathcal{G}\,]\ =\ 0\,,
\label{collective g relations}
\end{equation}
where $\pi_1^{(0,0)}\,, \pi_1^{(1,0)}\,, \pi_1^{(0,1)}$\,, and $\pi_1^{(1,1)}$
are the projection operators onto 
$\mathcal{H}_{\NSNS}\,, \mathcal{H}_{\RNS}\,, \mathcal{H}_{\RR}$\,, and
$\mathcal{H}_{\RR}$ components of the Hilbert space $\mathcal{H}$\,, respectively.
It is also useful to define
\begin{equation}
 \mathcal{X}\ =\ \pi_1^{(0,*)}\id + \pi_1^{(1,*)}X\,,\qquad
 \bar{\mathcal{X}}\ =\ \pi_1^{(*,0)}\id + \pi_1^{(*,1)}\bar{X}\,,
\end{equation}
with $\pi_1^{(r,*)}=\pi_1^{(r,0)}+\pi_1^{(r,1)}$
and $\pi_1^{(*,r)}=\pi_1^{(0,r)}+\pi_1^{(1,r)}$ $(r=0,1)$\,; then we can write
$\mathcal{G}=\mathcal{X}\bar{\mathcal{X}}$\,.
Note that $\mathcal{G}^{-1}$ is the inverse of $\mathcal{G}$ in this sense.
Then we can define the projection operator 
$\mathcal{P}_{\mathcal{G}}=\mathcal{G}\mathcal{G}^{-1}$ and collectively write
the extra constraints of Eqs.~(\ref{restrict RNS}) and (\ref{constraint RR}) as
\begin{equation}
\mathcal{P}_{\mathcal{G}}\Phi=\Phi\,.  
\label{restricted}
\end{equation} 
We call the Hilbert space restricted by the constraints in Eqs.~(\ref{restrict closed}) and
(\ref{restricted}) the restricted Hilbert space, 
or frequently simply restricted space, in this paper. 
On the restricted Hilbert space,
the BRST operator acts consistently: 
\begin{equation}
\mathcal{P}_{\mathcal{G}}Q\mathcal{P}_{\mathcal{G}}=Q\mathcal{P}_{\mathcal{G}}\,.
\label{BRST in restricted}
\end{equation}

A natural symplectic form in the restricted Hilbert space is defined
as follows.
First, the symplectic form in the space restricted by the closed string constraints
of Eq.~(\ref{restrict closed}) is defined by using the BPZ inner-product as
\begin{equation}
 \omega_s(\Phi_1\,,\Phi_2)\ =\ (-1)^{|\Phi_1|}\langle\Phi_1|c_0^-|\Phi_2\rangle\,,
\label{symplectic closed}
\end{equation}
where $\langle\Phi|$ is the BPZ conjugate of
 $|\Phi\rangle$\,.
The symbol $|\Phi|$ denotes the Grassmann property of the string field $\Phi$\,: 
$|\Phi|=0$ or $1$ if $\Phi$ is Grassmann even or odd, respectively.
For later use, we also define the symplectic forms 
$\omega_m$\,, $\omega_{\bar{m}}$\,, and $\omega_l$ in the Hilbert 
spaces
\begin{equation}
 \mathcal{H}_m\ =\ \mathcal{H}_{large}\otimes\bar{\mathcal{H}}\,,\qquad
 \mathcal{H}_{\bar{m}}\ =\ \mathcal{H}\otimes\bar{\mathcal{H}}_{large}\,,\qquad
 \mathcal{H}_l\ =\ \mathcal{H}_{large}\otimes\bar{\mathcal{H}}_{large}\,,
\label{symplectic forms}
\end{equation}
by
\begin{equation}
\omega_i(\varphi_1\,,\varphi_2)\ =\ 
(-1)^{|\varphi_1|}{}_i\langle\varphi_1|c_0^-|\varphi_2\rangle_i\,,\qquad
i\ =\ m\,,\bar{m}\,,l
\end{equation}
where ${}_i\langle\varphi_1|$ and ${}_i\langle\varphi_2|$ are 
the BPZ conjugates of $|\varphi_1\rangle_i$ and $|\varphi_2\rangle_i$
in $\mathcal{H}_i$ $(i=m,\bar{m},l)$\,, respectively.
The symplectic form $\omega_s$\,, $\omega_m$ and $\omega_{\bar{m}}$
are related to $\omega_l$ as
\begin{subequations} 
 \begin{align}
 \omega_s(\Phi_1\,,\Phi_2)\ =&\ \omega_l(\xi\bar{\xi}\Phi_1\,,\Phi_2)\,,\\
 \omega_m(\Phi_1\,,\Phi_2)\ =&\ - \omega_l(\bar{\xi}\Phi_1\,,\Phi_2)\,,\\
 \omega_{\bar{m}}(\Phi_1\,,\Phi_2)\ =&\ \omega_l(\xi\Phi_1\,,\Phi_2)\,,
\end{align}
\end{subequations}
for $\Phi_1,\ \Phi_2\in\mathcal{H}_i$ $(i=s,m,\bar{m})$\,.
Then, a natural symplectic form in the restricted Hilbert space is defined by
\begin{align}
 \Omega(\Phi_1\,,\Phi_2)\ =&\ \omega_s(\Phi_1\,,\mathcal{G}^{-1}\Phi_2)
\nonumber\\
=&\ \omega_s(\Phi_{1\NSNS}\,,\Phi_{2\NSNS})
+ \omega_s(\Phi_{1\RNS}\,,Y\Phi_{2\RNS})
\nonumber\\
&
+ \omega_s(\Phi_{1\NSR}\,,\bar{Y}\Phi_{2\NSR})
+ \omega_s(\Phi_{1\RR}\,,\mathcal{Y}\bar{\mathcal{Y}}\Phi_{2\RR})\,.
\label{symplectic Omega}
\end{align}
This has the non-degenerate cross-diagonal form common in each sector
\begin{align}
 \Omega(\Phi_1\,, \Phi_2)\ 
=\ \sum_{I}\Big(
\llangle\phi_{1I}|\psi_{2I}\rrangle
+ \llangle\psi_{1I}|\phi_{2I}\rrangle\Big)\,,
\label{Omega all}
\end{align}
after integrating out (or carrying out the inner product of)
the ghost zero modes,\footnote{
The expansion with respect to the ghost zero mode is summarized
in Appendix \ref{ghost zero modes}.} 
where the index $I$ runs over each component, 
NS-NS, R-NS, NS-R, and R-R.
The fields $\phi_I$ and $\psi_I$
are subcomponents of each component $\Phi_I$
expanded with respect to the ghost zero modes as in 
Eqs.~(\ref{restrict NSNS}), (\ref{restricted RNS and NSR}), 
and (\ref{restricted RR}).
It should be noted that the $\psi_{\RR}$ in Eq.~(\ref{restricted RR}) 
is defined so that the non-locality 
of $\mathcal{Y}$ and $\bar{\mathcal{Y}}$ in the R-R sector 
disappears in the symplectic form in Eq.~(\ref{Omega all}).
In the following, we will see that this cross-diagonal form of 
the symplectic form $\Omega$ in the restricted space
provides a free field theory which can properly 
be quantized via the BV formalism.

\subsection{Free field theory}\label{free theory}

Using the symplectic form $\Omega$ in the restricted Hilbert space,
the free field action and gauge transformation 
of the type II superstring field theory are given by
\begin{equation}
 S_0\ =\ \frac{1}{2}\Omega(\Phi\,, Q\Phi)\,,\qquad
 \delta\Phi\ =\ Q\Lambda\,,
\label{classical action and gauge}
\end{equation}
where the gauge parameter
also has four components,
\begin{equation}
 \Lambda\ =\  \Lambda_{\NSNS} +  \Lambda_{\RNS}
+  \Lambda_{\NSR} + \Lambda_{\RR}\,,
\end{equation}
is Grassmann odd, and has ghost number 1.
The equation of motion 
\begin{equation}
0\ =\ \mathcal{G}^{-1}Q\Phi\,,
\end{equation}
derived from Eq.~(\ref{classical action and gauge})
can be written by using Eq.~(\ref{BRST in restricted}) as
\begin{equation}
0\ =\ \mathcal{G}^{-1}Q\mathcal{P}_{\mathcal{G}}\Phi\
=\ \mathcal{G}^{-1}\mathcal{P}_{\mathcal{G}}Q\mathcal{P}_{\mathcal{G}}\Phi\,.
\end{equation}
Then, by multiplying by $\mathcal{G}$\,, we have
\begin{equation}
0\ =\ \mathcal{P}_{\mathcal{G}}Q\mathcal{P}_{\mathcal{G}}\Phi\ =\ 
Q\mathcal{P}_{\mathcal{G}}\Phi\ =\ Q\Phi\,,
\label{classical eom}
\end{equation}
thanks to Eq.~(\ref{collective g relations}).

The action in Eq.~(\ref{classical action and gauge}) has superficially the same form
as that of the bosonic string field theory, and its BV quantization can be 
carried out in a similar way \cite{Bochicchio:1986bd}.
The master action $\bd{S}_0$ 
can simply 
be given by removing the ghost number restriction on $\Phi$ in the classical action:
\begin{equation}
 \bd{S}_{0}^{}\ =\ \frac{1}{2}\,\Omega(\boldsymbol{\Phi}\,,
Q\boldsymbol{\Phi})\,,
\label{master action}
\end{equation}
where $\boldsymbol{\Phi}=\sum_{g=-\infty}^\infty \Phi^{(g)}$\,. Each
$\Phi^{(g)}$ is the string field with the ghost number $g$\,. 
The component $\Phi^{(2)}$ is equal to the classical string field $\Phi$\,,
and the others are the space-time ghosts, anti-ghosts, 
and corresponding anti-fields.
The BRST transformation, which keeps the master action in Eq.~(\ref{master action}) 
invariant, is obtained by replacing the parameter $\Lambda$ 
in the gauge transformation of Eq.~(\ref{classical action and gauge}) 
by the field $\boldsymbol{\Phi}$ as
\begin{equation}
\delta_{B0}\boldsymbol{\Phi}\ =\ Q\boldsymbol{\Phi}\,. 
\label{BRST tf}
\end{equation}
It is easy to show that the master action in Eq.~(\ref{master action}) actually satisfies
the BV master equation.
Using the fact that the symplectic form $\Omega$ has 
the cross-diagonal form of Eq.~(\ref{Omega all}),
an arbitrary variation of the master action can be written as
\begin{align}
\delta \bd{S}_0\ 
=&\ \Omega(\delta\boldsymbol{\Phi}\,,\delta_{B0}\boldsymbol{\Phi})
\nonumber\\
=&\ \sum_I\Big(\llangle\delta\boldsymbol{\phi}_I|\delta_{B0}\boldsymbol{\psi}_I\rrangle
+ \llangle\delta\boldsymbol{\psi}_I|\delta_{B0}\boldsymbol{\phi}_I\rrangle
\Big)\,,
\end{align}
and thus we have
\begin{equation}
 \frac{\partial \bd{S}_0}{\partial\boldsymbol{\phi}_I}\ 
=\ \delta_{B0}\boldsymbol{\psi}_I\,,\qquad
 \frac{\partial \bd{S}_0}{\partial\boldsymbol{\psi}_I}\ 
=\ \delta_{B0}\boldsymbol{\phi}_I\,.
\end{equation}
The BRST invariance of the action implies that
the classical BV master equation holds:
\begin{align}
 0\ =&\ 
\sum_I\Bigg(\frac{\partial \bd{S}_0}{\partial\boldsymbol{\phi}_I}\delta_{B0}\boldsymbol{\phi}_I
+ \frac{\partial \bd{S}_0}{\partial\boldsymbol{\psi}_I}\delta_{B0}\boldsymbol{\psi}_I\Bigg)
\nonumber\\
=&\ 2\sum_I\Bigg(
\frac{\partial \bd{S}_0}{\partial\boldsymbol{\phi}_I}
\frac{\partial \bd{S}_0}{\partial\boldsymbol{\psi}_I}\Bigg)\,.
\end{align}
The components $\boldsymbol{\phi}_I$ and $\boldsymbol{\psi}_I$ are 
identified with the fields and anti-fields
in the gauge-fixed basis in the BV formulation \cite{Kohriki:2012pp},
respectively.\footnote{
Here, since the field $\boldsymbol{\phi}_I$ is Grassmann even, the anti-field $\boldsymbol{\psi}_I$ 
must be Grassmann odd. We can show that this is actually true
for the GSO projected string field, which we implicitly assume \cite{Kohriki:2012pp}.
} 
The gauge-fixed action in the Siegel gauge 
is obtained by setting $\boldsymbol{\psi}_I=0$\,.

\subsection{R-R action}

Before incorporating the interactions, let us examine 
the action of the R-R sector,  
\begin{equation}
S_0^{\RR}\ 
 =\ \frac{1}{2}\omega_s(\Phi_{\RR}\,, \mathcal{Y}\bar{\mathcal{Y}} Q\Phi_{\RR})\,,
\label{action RR} 
\end{equation}
in more detail since it is characteristic 
of type II superstring field theory.
For simplicity 
we take the Siegel gauge $|\boldsymbol{\psi}_{\RR}\rrangle=0$
in this discussion.
After integrating out the ghost zero modes,
the master action of Eq.~(\ref{master action}) and 
the BRST transformation of Eq.~(\ref{BRST tf})
in the R-R sector become
\begin{equation}
 \bd{S}_0^{\RR}\ =\ 
\llangle\boldsymbol{\phi}_{\RR}|\frac{2G\bar{G}}{L_0^+}|\boldsymbol{\phi}_{\RR}\rrangle\,,
\qquad 
 \delta_{B0}|\boldsymbol{\phi}_{\RR}\rrangle\ =\ 
\tilde{Q}|\boldsymbol{\phi}_{\RR}\rrangle\,.
\label{action RR phi psi}
\end{equation}
Although this action is non-local, the propagator
\begin{equation}
-\frac{b_0^+b_0^-\delta(\bar{\beta}_0)\delta(\beta_0)G\bar{G}\delta(L_0^-)}{L_0^+}
\end{equation}
agrees with that obtained in the first-quantized formulation \cite{Witten:2012bh}.
%
%
%
%

If one wants to avoid the non-local action,
one can replace it with the Sen-like action as an alternative by introducing an extra 
Grassmann even string field $\tilde{\Phi}_{\RR}$\,, which 
is restricted by the closed string constraints in Eq.~(\ref{restrict closed})
and has ghost number 2 and picture number $-3/2$\,.
Then the alternative action is given by
\begin{align}
 \tilde{S}_0^{\RR}\ =&\ - \frac{1}{2}\omega_s(\tilde{\Phi}_{\RR}\,,X\bar{X}Q\tilde{\Phi}_{\RR})
+ \omega_s(\tilde{\Phi}_{\RR}\,,Q\Phi_{\RR})
\nonumber\\
=&\ - \frac{1}{2}\Omega(X\bar{X}\tilde{\Phi}_{\RR}\,, 
QX\bar{X}\tilde{\Phi}_{\RR})
+ \Omega(X\bar{X}\tilde{\Phi}_{\RR}\,,Q\Phi_{\RR})\,.
\label{Sen like action}
\end{align}
The difference from Sen's original action, however, is that
we can rewrite it using the method of completing the square as
\begin{equation}
 \tilde{S}_0^{\RR}\ =\ - \frac{1}{2}\Omega(X\bar{X}\tilde{\Phi}'_{\RR}\,, 
QX\bar{X}\tilde{\Phi}'_{\RR})
+\frac{1}{2}\Omega(\Phi_{\RR}\,,Q\Phi_{\RR})\,,
\end{equation}
with $\tilde{\Phi}'_{\RR}=\tilde{\Phi}_{\RR}-\mathcal{Y}\bar{\mathcal{Y}}\Phi_{\RR}$\,,
thanks to the constraint in Eq.~(\ref{constraint RR}), where the equivalence is obvious.
Since $\tilde{\Phi}_{\RR}$ appears only in the form of $X\bar{X}\tilde{\Phi}_{\RR}$
in the action of Eq.~(\ref{Sen like action}), we can restrict $\tilde{\Phi}_{\RR}$ 
by the condition
\begin{equation}
 \mathcal{Y}X\tilde{\Phi}_{\RR}\ =\ 
\bar{\mathcal{Y}}\bar{X}\tilde{\Phi}_{\RR}\ =\ \tilde{\Phi}_{\RR}\,,
\label{constraint tilde RR}
\end{equation}
which is dual to the constraint in Eq.~(\ref{constraint RR}) on $\Phi_{\RR}$
and restricts $\tilde{\Phi}_{\RR}$ to the form of
\begin{equation}
 \tilde{\Phi}_{\RR}\ =\ \tilde{\phi}_{\RR} - c_0^+\tilde{\psi}_{\RR}\,.
\end{equation}
The Sen-like master action in the generalized Siegel gauge 
$\boldsymbol{\psi}_{\RR}=\boldsymbol{\tilde{\psi}}_{\RR}=0$ then becomes
\begin{align}
 \tilde{\bd{S}}_0^{\RR}\ =&\
  \frac{1}{2}\llangle\boldsymbol{\tilde{\phi}}_{\RR}|L_0^+G\bar{G}|\boldsymbol{\tilde{\phi}}_{\RR}\rrangle
- \llangle\boldsymbol{\tilde{\phi}}_{\RR}|L_0^+|\boldsymbol{\phi}_{\RR}\rrangle
\nonumber\\
=&\ \frac{1}{2}\llangle\boldsymbol{\tilde{\phi}}_{\RR}'|L_0^+G\bar{G}|\boldsymbol{\tilde{\phi}}_{\RR}'\rrangle
+ \frac{1}{2}\llangle\boldsymbol{\phi}_{\RR}|\frac{4G\bar{G}}{L_0^+}|\boldsymbol{\phi}_{\RR}\rrangle\,,
\end{align}
with $|\boldsymbol{\tilde{\phi}}_{\RR}'\rrangle=|\boldsymbol{\tilde{\phi}}_{\RR}\rrangle
+4\frac{G\bar{G}}{(L_0^+)^2}|\boldsymbol{\phi}_{\RR}\rrangle$
after integrating out the ghost zero modes. 
Although the extra sector is a higher-derivative theory,
it stays free and is decoupled from the physical sector if
the interaction part of the action does not include $\tilde{\Phi}_{\RR}$\,.

\subsection{Including interactions}\label{interaction}

The interactions of type II superstring field theory are described 
by the multi-string products,
 \begin{equation}
  L_n(\Phi_1\,,\cdots\,,\Phi_n)\,,\qquad (n\ge1)\,,
\label{Ln general}
 \end{equation}
which make a string field from $n$ string fields $\Phi_1,\cdots,\Phi_n$\,.
They are graded symmetric under interchange of the $n$ string fields,
and must carry proper ghost number and picture number.
In addition, since the type II superstring field in this formulation
is in the restricted small Hilbert space, 
the outputs of the string products must also satisfy the constraint
in Eq.~(\ref{restricted}): 
\begin{equation}
\mathcal{P}_{\mathcal{G}}L_n(\Phi_1\,,\cdots\,,\Phi_n)\ =\ 
L_n(\Phi_1\,,\cdots\,,\Phi_n)\,.
\end{equation}
By using these string products, the action of
type II superstring field theory is given by
\begin{equation}
 S\ =\ \sum_{n=0}^\infty\frac{1}{(n+2)!}\Omega(\Phi\,,L_{n+1}(\underbrace{\Phi\,,\cdots\,,\Phi}_{n+1}))\,,
\label{action}
\end{equation}
where $L_1$ is identified with the BRST operator: $L_1=Q$\,. 
The action in Eq.~(\ref{action}) is invariant 
under the gauge transformation
\begin{equation}
\delta \Phi\ =\ \sum_{n=0}^\infty\frac{1}{n!}L_{n+1}(\underbrace{\Phi\,,\cdots\,,\Phi}_{n}\,,\Lambda)\,,
\end{equation}
if the string products satisfy the $L_\infty$ relations
\begin{equation}
 \sum_\sigma\sum_{m=1}^n(-1)^{\epsilon(\sigma)}\frac{1}{m!(n-m)!}
L_{n-m+1}(L_m(\Phi_{\sigma(1)}\,,\cdots\,,\Phi_{\sigma(m)})\,,\Phi_{\sigma(m+1)}\,,\cdots\,,\Phi_{\sigma(n)})\
=\ 0
\label{L infinity}
\end{equation}
and cyclicity with respect to the symplectic form $\Omega$\,:
\begin{align}
 \Omega(\Phi_1\,, L_n(\Phi_2\,,\cdots\,,\Phi_{n+1}))\ =&\
-(-1)^{|\Phi_1|}\Omega(L_n(\Phi_1\,,\cdots\,,\Phi_n)\,, \Phi_{n+1})
\,.
\label{cyclicity L}
\end{align}
Here, the symbol $\sigma$ denotes the permutation from
$\{1,\cdots,n\}$ to $\{\sigma(1),\cdots,\sigma(n)\}$\,, and 
$\epsilon(\sigma)$ is the sign factor of the permutation of the string fields
from $\{\Phi_1,\cdots,\Phi_n\}$ to $\{\Phi_{\sigma(1)}\,\cdots,\Phi_{\sigma(n)}\}$\,.
If the set of string products $\{L_n\}$ satisfies these conditions, it is called
the string products with the cyclic $L_\infty$ structure or simply
the cyclic $L_\infty$ algebra. 
The problem is how to construct such an $L_\infty$ algebra.

\sectiono{Construction of string products with $L_\infty$ structure}\label{cyclic products}

Let us construct a set of string products realizing a cyclic $L_\infty$ algebra.
We use a coalgebraic representation handling an infinite number of string products
in the $L_\infty$ algebra collectively. We follow the notation and convention
in Ref.~\cite{Kunitomo:2019glq}.

\subsection{Cyclicity, Ramond numbers and picture number deficit}\label{cyclicity}

String products describing the interaction of type II superstrings must
have a proper ghost number and picture number. 
Since the ghost number structure is the same as that of the bosonic closed string
field theory, here it is enough to consider the picture number that 
the string products should have. 
Denote a coderivation corresponding to an $(n+2)$-string product $(n\ge0)$
with picture number $(p,p')$ as $\bd{B}^{(p,p')}_{n+2}$\,.
In order to describe the type II superstring interaction, the output string state
must have the same picture number as that of the type II superstring field:
the picture number of its NS-NS, R-NS, NS-R, and R-R components must be equal
to $(-1,-1)$\,, $(-1/2,-1)$\,, $(-1,-1/2)$\,, and $(-1/2,-1,2)$\,, respectively.
The string products are also characterized by their Ramond and cyclic Ramond 
number defined by
\begin{equation}
 \begin{pmatrix} 
  \textrm{Ramond}\\
 \textrm{cyclic Ramond}
 \end{pmatrix}
\textrm{number}\ =\ \#\ \textrm{of Ramond inputs} \mp \#\ \textrm{of Ramond outputs}\,,
\end{equation}
which are also assigned for each of the left- and right-moving sectors.
Since we can consider each sector separately let us first consider
the left-moving sector.
Suppose that $2r$ of $n+2$ inputs are the R states in the left-moving sector. 
If we assume conservation of the space-time fermion number
then the output must be the NS state, and thus
\begin{subequations} \label{picture numbers}
\begin{equation}
 \left(-\frac{1}{2}\right)\times 2r + (-1)\times(n+2-2r) + p\ =\ -1
\end{equation}
from the picture number conservation. Such a string product is characterized
by the cyclic Ramond number $2r$ and the Ramond number $2r$\,.
If $2r+1$ of the inputs are the R states,
the output is the R state and
\begin{equation}
 \left(-\frac{1}{2}\right)\times(2r+1)+(-1)\times(n+1-2r) + p\ =\ -\frac{1}{2}\,,
\end{equation}
\end{subequations}
which is the case characterized by the cyclic Ramond number $2r+2$
and the Ramond number $2r$\,. Both of these equations in Eq.~(\ref{picture numbers}) 
can be solved as $n=p+r-1$\,. 
After repeating the same consideration for the right-moving sector,
we can find that a candidate coderivation describing the type II superstring 
interaction is the one respecting the Ramond number:\footnote{
We take the convention that the quantity with the (cyclic) Ramond number 
out of range is identically equal to zero as in Ref.~\cite{Kunitomo:2019glq}.}
\begin{align}
 \sum_{p,r,p',r'=0}^\infty\delta_{p+r,p'+r'}&
\Big(\pi_1^{(0,0)}\bd{B}_{p+r+1}^{(p,p')}\big{|}^{(2r,2r')}_{(2r,2r')}
+ \pi_1^{(1,0)}\bd{B}_{p+r+1}^{(p,p')}\big{|}^{(2r+2,2r')}_{(2r,2r')}
\nonumber\\
&\
+ \pi_1^{(0,1)}\bd{B}_{p+r+1}^{(p,p')}\big{|}^{(2r,2r'+2)}_{(2r,2r')}
+ \pi_1^{(1,1)}\bd{B}_{p+r+1}^{(p,p')}\big{|}^{(2r+2,2r'+2)}_{(2r,2r')}
\Big)
\nonumber\\
=&\ \sum_{p,r,p',r'=0}^\infty \delta_{p+r,p'+r'}
\pi_1\bd{B}_{p+r+1}^{(p,p')}\big{|}_{(2r,2r')}\,,
\end{align}
with 
%
$\bd{B}_1\equiv0$\,, 
%
which we call the string products with no picture number deficit.
However, 
this is not suitable for considering the cyclicity
since the Ramond number is not invariant under the cyclic permutation as in 
Eq.~(\ref{cyclicity L}).
So, instead we consider the string products,
\begin{align}
 \pi_1\bd{B}\ \equiv&
\sum_{p,,r,r'=0}^\infty\pi_1\bd{B}_{p+r+1}^{(p,p')}\big{|}^{(2r,2r')}
\nonumber\\
=&\ 
\sum_{p,r,r'=0}^\infty 
\Big(\pi_1^{(0,0)}\bd{B}_{p+r+1}^{(p,p')}\big{|}^{(2r,2r')}_{(2r,2r')}
+ \pi_1^{(1,0)}\bd{B}_{p+r+1}^{(p,p')}\big{|}^{(2r,2r')}_{(2r-2,2r')}
\nonumber\\
&\hspace{20mm}
+ \pi_1^{(0,1)}\bd{B}_{p+r+1}^{(p,p')}\big{|}^{(2r,2r')}_{(2r,2r'-2)}
+ \pi_1^{(1,1)}\bd{B}_{p+r+1}^{(p,p')}\big{|}^{(2r,2r')}_{(2r-2,2r'-2)}
\Big)\,,
\label{B cyclic Ramond}
\end{align}
respecting the cyclic Ramond number that is invariant under the permutation.
While it becomes easy to consider the cyclicity,
this combination of string products $\bd{B}$ cannot be used as it is
since its NS-NS, R-NS, NS-R, and R-R components
have picture number deficit $(0,0)$\,, $(1,0)$\,, $(0,1)$\,, 
and $(1,1)$\,, respectively.
In the heterotic string field theory, similar string products
can naturally appear in a non-linear extension of the combination 
of the operator (one-string product) $\bd{Q}-\bd{\eta}$ 
\cite{Kunitomo:2019glq}. In type II superstring theory, however, the analogous
combination $\bd{Q}-\bd{\eta}-\bar{\bd{\eta}}$ has no counterpart with 
picture number deficit $(1,1)$\,, and thus we cannot directly extend the prescription
in Ref.~\cite{Kunitomo:2019glq} to construct the required $L_\infty$ algebra. 
We take an alternative way that is a generalization of the asymmetric construction 
used in Ref.~\cite{Erler:2014eba} to give those restricted into the NS-NS sector.

\subsection{Construction of string products}\label{string products}

The prescription we propose is simply repeating that used in heterotic
string field theory twice:
the first time is for getting the correct structure of the left-moving
sector by inserting $X$ and/or $\xi$ in the bosonic string products, which we assume 
to be known \cite{Saadi:1989tb,Kugo:1989aa,Zwiebach:1992ie}, 
and the second time is for getting 
the correct structure of the right-moving sector by inserting $\bar{X}$ and/or $\bar{\xi}$
in the (heterotic) string products obtained in the first step\,.

We start from the combined coderivation
\begin{equation}
 \bd{D} - \bd{C}\ =\ \bd{Q} - \bd{\eta} + \bd{B}\,,
\end{equation}
with
\begin{equation}
\bd{B}\ =\ \sum_{p,r=0}^\infty \bd{B}^{(p)}_{p+r+1}\big{|}^{2r}\,,
\end{equation}
where $p$ and $2r$ are the picture and cyclic Ramond numbers of 
the left-moving sector, respectively. 
This can be decomposed to $\bd{D}$ and $\bd{C}$ by 
picture number deficit as
\begin{align}
 \pi_1\bd{D}\ =&\ 
\pi_1\bd{Q} + \sum_{p,r=0}^\infty \pi_1\bd{B}^{(p)}_{p+r+1}\big{|}^{2r}_{2r}\
=\ \pi_1\bd{Q} + \pi_1^{(0,*)}\bd{B}\,,\\
\pi_1\bd{C}\ =&\ 
\pi_1\bd{\eta} - \sum_{p,r=0}^\infty \pi_1\bd{B}^{(p)}_{p+r+1}\big{|}^{2r}_{2r-2}\
=\ \pi_1\bd{\eta} - \pi_1^{(1,*)}\bd{B}\,.
\label{left decomposition}
\end{align}
Suppose $\bd{B}$ has zero right-moving picture number and is independent
of the right-moving Ramond and cyclic Ramond numbers. 
The left-moving picture number deficit 
of $\bd{D}$ is equal to zero and that of $\bd{C}$ is equal to one.
As was shown in Ref.~\cite{Kunitomo:2019glq}, 
the $L_\infty$ relation for the coderivation $\bd{D}-\bd{C}$\,,
\begin{equation}
 [\,\bd{D}-\bd{C}\,,\bd{D}-\bd{C}\,]\ =\ 0\,,
\end{equation}
following from the equations
\begin{subequations} \label{eq left}
 \begin{align}
&\ [\,\bd{Q}\,,\bd{B}(s,t)\,] + \frac{1}{2}[\,\bd{B}(s,t)\,,\bd{B}(s,t)\,]^1
+\frac{s}{2}[\,\bd{B}(s,t)\,,\bd{B}(s,t)\,]^2\ =\ 0\,,\\
&\ [\,\bd{\eta}\,,\bd{B}(s,t)\,] - \frac{t}{2}[\,\bd{B}(s,t)\,,\bd{B}(s,t)\,]^2\ =\ 0\,,
 \end{align}
\end{subequations}
for the generating function 
\begin{align}
 \bd{B}(s,t)\ =&\ \sum_{m,n,r=0}^\infty s^m t^n \bd{B}^{(n)}_{m+n+r+1}\big{|}^{2r}\
=\ \sum_{n=0}^\infty t^n\bd{B}^{(n)}(s)\,,
\end{align}
from which we obtain the required string products as $\bd{B}=\bd{B}(0,1)$\,.
The operations $[\bd{l},\bd{l}']^1$ amd $[\bd{l},\bd{l}']^2$ in Eq.~(\ref{eq left})
are defined by splitting the commutator
into the pieces with definite left-moving cyclic Ramond number: if $\bd{l}=\sum_r\bd{l}|^{2r}$ and
$\bd{l}'=\sum_{r'}\bd{l}'|^{2r'}$\,, then
\begin{equation}
[\,\bd{l}\,,\bd{l}'\,]^1\ =\ \sum_{r,r'}\,[\,\bd{l}|^{2r}\,,\bd{l}'|^{2r'}\,]|^{2r+2r'}\,,
\qquad
[\,\bd{l}\,,\bd{l}'\,]^2\ =\ \sum_{r,r'}\,[\,\bd{l}|^{2r}\,,\bd{l}'|^{2r'}\,]|^{2r+2r'-2}\,.
\label{operations}
\end{equation}
We can show that $[\bd{l},\bd{l}']=[\bd{l},\bd{l}']^1+[\bd{l},\bd{l}']^2$\,.
It was also shown in Ref.~\cite{Kunitomo:2019glq} that the equations in 
Eq.~(\ref{eq left}) are satisfied
if we postulate the differential equations
\begin{subequations} \label{diff left}
 \begin{align}
 \partial_t\bd{B}(s,t)\ =&\ [\,\bd{Q}\,,\bd{\lambda}(s,t)\,] +
 [\,\bd{B}(s,t)\,,\bd{\lambda}(s,t)\,]^1 + s\,  [\,\bd{B}(s,t)\,,\bd{\lambda}(s,t)\,]^2\,,\\
[\,\bd{\eta}\,,\bd{\lambda}(s,t)\,]\ =&\ \partial_s\bd{B}(s,t) 
+ t\, [\,\bd{B}(s,t)\,,\bd{\lambda}(s,t)\,]^2
\end{align}
\end{subequations}
by introducing (a generating function of) the gauge products represented by a degree-even coderivation
\begin{equation}
 \bd{\lambda}(s,t)\ =\ \sum_{m,n,r=0}^\infty s^m t^n
  \bd{\lambda}^{(n+1)}_{m+n+r+2}\big{|}^{2r}\
=\ \sum_{n=0}^\infty t^n\bd{\lambda}^{(n+1)}(s)\,.
\label{gf lambda}
\end{equation}
The differential equations in Eq.~(\ref{diff left}) can be recursively solved as
\begin{subequations} \label{lambda B}
 \begin{align}
\bd{\lambda}^{(n+1)}(s)\ =&\ \xi\circ\left(\partial_s\bd{B}^{(n)}(s)
+ \sum_{n'=0}^{n-1}[\,\bd{B}^{(n-n')}(s)\,,\bd{\lambda}^{(n'+1)}(s)\,]^2\right)\,,
\label{lambda}\\
(n+1)\bd{B}^{(n+1)}(s)\ =&\ [\,\bd{Q}\,,\bd{\lambda}^{(n+1)}(s)\,]
\nonumber\\
&\
+ \sum_{n'=0}^n[\,\bd{B}^{(n-n')}(s)\,,\bd{\lambda}^{(n'+1)}(s)\,]^1
+ \sum_{n'=0}^ns\,[\,\bd{B}^{(n-n')}(s)\,,\bd{\lambda}^{(n'+1)}(s)\,]^2\,,
\label{B}
\end{align}
\end{subequations}
with the initial condition 
\begin{equation}
 \bd{B}^{(0)}(s)=\bd{L}_B(s)\,,
\end{equation}
given by using the interacting part of the bosonic products
(string products with no non-zero picture number operator insertion)\cite{Kunitomo:2019glq}:
\begin{equation}
\bd{L}_B(s)\ =\ \sum_{m,r=0}^\infty s^m \bd{L}^B_{m+r+1}\big{|}^{2r}\,,\qquad
(\bd{L}^B_1\big{|}^0\equiv0)\,. 
\end{equation}
The operation $\xi\circ$ in Eq.~(\ref{lambda}) is defined as that inserting $\xi$ cyclically.
Then, by construction, all the $\bd{B}(s,t)$ and $\bd{\lambda}(s,t)$ are cyclic with 
respect to the symplectic form $\omega_m$\,.
They provide a cyclic $L_\infty$ algebra $(\mathcal{H}_m,\omega_m,\bd{D}-\bd{C})$\,.
After decomposing this combined $L_\infty$ algebra into $\bd{D}$ and $\bd{C}$\,,
we can obtain a \textit{heterotic} $L_\infty$ algebra
in the small Hilbert space, satisfying 
$[\bd{\eta},\bd{L}_H]=[\bar{\bd{\eta}},\bd{L}_H]=0$\,,
by similarity transformation generated by the cohomomorphism
\begin{equation}
 \pi_1\hat{\bd{F}}^{-1}\ =\ \pi_1\idSH - \Xi\pi_1^{(1,*)}\bd{B}
\end{equation}
as
\begin{equation}
 \pi_1\bd{L}_H\ \equiv\ 
\pi_1\hat{\bd{F}}^{-1}\bd{D}\hat{\bd{F}}
=\ \pi_1\bd{Q} + \mathcal{X}\bd{b}_H\,,
\label{heterotic L}
\end{equation}
with $\bd{b}_H=\bd{B}\hat{\bd{F}}$\,.
This $\bd{L}_H$ has the required picture number structure for the left-moving sector
but the right-moving picture number is still equal to zero:
\begin{align}
 \bd{L}_H\ =&\ \bd{Q} + \sum_{p,r=0}^\infty (\bd{L}_H)_{p+r+1}^{(p,0)}\big{|}_{2r}
=\ \bd{Q} + \sum_{p,r=0}^\infty \mathcal{X}(\bd{b}_H)_{p+r+1}^{(p,0)}\big{|}^{2r}\,.
\end{align}
Here, the subscript or superscript $2r$ after the vertical line is the left-moving Ramond 
or cyclic Ramond number, respectively. 
It is easy to see that $\bd{b}_H$ is cyclic with respect to $\omega_m$ 
in the same way as in Ref.~\cite{Kunitomo:2019glq}.

We repeat the same procedure for the right-moving sector.
Let us consider the combined coderivation
\begin{equation}
 \bar{\bd{D}} - \bar{\bd{C}}\ =\ \bd{Q} - \bar{\bd{\eta}} + \bar{\bd{B}}\,,
\end{equation}
with
\begin{equation}
 \bar{\bd{B}}\ =\ \sum_{\bar{p},\bar{r}=0}^\infty 
\bar{\bd{B}}^{(\bar{p})}_{\bar{p}+\bar{r}+1}\big{|}^{2\bar{r}}\,,
\label{bar B}
\end{equation}
which can be decomposed by the right-moving picture number deficit as
\begin{align}
 \pi_1\bar{\bd{D}}\ =&\ 
\pi_1\bd{Q} + \sum_{\bar{p},\bar{r}=0}^\infty 
\pi_1\bar{\bd{B}}^{(\bar{p})}_{\bar{p}+\bar{r}+1}|^{2\bar{r}}_{2\bar{r}}\
=\ \pi_1\bd{Q} + \pi_1^{(*,0)}\bar{\bd{B}}\,,\\
\pi_1\bar{\bd{C}}\ =&\ 
\pi_1\bar{\bd{\eta}} - \sum_{\bar{p},\bar{r}=0}^\infty 
\pi_1\bar{\bd{B}}^{(\bar{p})}_{\bar{p}+\bar{r}+1}|^{2\bar{r}}_{2\bar{r}-2}\
=\ \pi_1\bar{\bd{\eta}} - \pi_1^{(*,1)}\bar{\bd{B}}\,.
\end{align}
It is noted that only the right-moving quantum numbers,
the picture, Ramond and cyclic Ramond numbers, 
are specified. 
Those of the left-moving sector are implicit but determined properly
in our construction below.
The $L_\infty$ relation of the coderivation $\bar{\bd{D}}-\bar{\bd{C}}$\,,
\begin{equation}
 [\,\bar{\bd{D}}-\bar{\bd{C}}\,,\bar{\bd{D}}-\bar{\bd{C}}\,]\ =\ 0\,,
\end{equation}
again follows from the equations
\begin{subequations}  \label{eq right}
\begin{align}
&\ [\,\bd{Q}\,,\bar{\bd{B}}(s,t)\,] + \frac{1}{2}[\,\bar{\bd{B}}(s,t)\,,\bar{\bd{B}}(s,t)\,]^{\bar{1}}
+\frac{s}{2}[\,\bar{\bd{B}}(s,t)\,,\bar{\bd{B}}(s,t)\,]^{\bar{2}}\ =\ 0\,,\\
&\ [\,\bar{\bd{\eta}}\,,\bar{\bd{B}}(s,t)\,] 
- \frac{t}{2}[\,\bar{\bd{B}}(s,t)\,,\bar{\bd{B}}(s,t)\,]^{\bar{2}}\ =\ 0\,,
\end{align}
\end{subequations}
for the generating function
\begin{equation}
 \bar{\bd{B}}(s,t)\ =\ \sum_{\bar{m},\bar{n},\bar{r}=0}^\infty 
s^{\bar{m}} t^{\bar{n}} \bar{\bd{B}}^{(\bar{n})}_{\bar{m}+\bar{n}+\bar{r}+1}\big{|}^{2\bar{r}}
=\ \sum_{\bar{n}=0}^\infty t^{\bar{n}}\bar{\bd{B}}^{(\bar{n})}(s)\,.
\end{equation}
The string products in Eq.~(\ref{bar B}) are obtained as $\bar{\bd{B}}=\bar{\bd{B}}(0,1)$\,.
The operations $[\bd{l},\bd{l}']^{\bar{1}}$ and $[\bd{l},\bd{l}']^{\bar{2}}$ are the right-moving
counterpart of Eq.~(\ref{operations}), which split the commutator into the pieces with a definite
right-moving cyclic Ramond number.
Equations (\ref{eq right}) are satisfied if we postulate the differential equations
\begin{subequations}  \label{diff right}
\begin{align}
 \partial_t\bar{\bd{B}}(s,t)\ =&\ [\,\bd{Q}\,,\bar{\bd{\lambda}}(s,t)\,] +
 [\,\bar{\bd{B}}(s,t)\,,\bar{\bd{\lambda}}(s,t)\,]^{\bar{1}}
+ s\,  [\,\bar{\bd{B}}(s,t)\,,\bar{\bd{\lambda}}(s,t)\,]^{\bar{2}}\,,
\label{eq right a}\\
[\,\bar{\bd{\eta}}\,,\bar{\bd{\lambda}}(s,t)\,]\ =&\ \partial_s\bar{\bd{B}}(s,t) 
+ t\, [\,\bar{\bd{B}}(s,t)\,,\bar{\bd{\lambda}}(s,t)\,]^{\bar{2}}
\label{eq right b}
\end{align}
\end{subequations}
by introducing (a generating function of) the gauge products represented 
by a degree-even coderivation
\begin{align}
 \bar{\bd{\lambda}}(s,t)\ =&\ \sum_{\bar{m},\bar{n},\bar{r}=0}^\infty 
s^{\bar{m}}
 t^{\bar{n}}\bar{\bd{\lambda}}^{(\bar{n}+1)}_{\bar{m}+\bar{n}+\bar{r}+2}\big{|}^{2\bar{r}}\
=\ \sum_{\bar{n}=0}^\infty t^{\bar{n}}\bar{\bd{\lambda}}^{(\bar{n}+1)}(s)\,.
\label{gf lambda bar}
\end{align}
This time we solve the differential equations in Eq.~(\ref{diff right}) by starting from
the initial condition $\bar{\bd{B}}^{(0)}(s)=\bd{L}_H(s)$
with
\begin{equation}
\bd{L}_H(s)\ =\ 
\sum_{\bar{m},\bar{r}=0}^\infty s^{\bar{m}}
(\bd{L}_H)_{\bar{m}+\bar{r}+1}\big{|}^{2\bar{r}}\,,
\end{equation}
or $\bar{\bd{b}}^{(0)}(s)=\bd{b}_H(s)$ with
$\bar{\bd{B}}^{(0)}(s)=\mathcal{X}\bar{\bd{b}}^{(0)}(s)$
and $\bd{L}_H(s)=\mathcal{X}\bd{b}_H(s)$\,.
Solving Eq.~(\ref{eq right b}) explicitly, we first obtain
\begin{equation}
 \bar{\bd{\lambda}}^{(1)}(s)\ =\ \bar{\xi}\circ\partial_s\bd{L}_H(s)
=\
\mathcal{X}\left(\bar{\xi}\circ\partial_s\bd{b}_H\right)\
\equiv\ \mathcal{X}\bd{\mu}^{(1)}(s)\,,
\label{lambda bar 1}
\end{equation}
and then, from Eq.~(\ref{eq right a}),
\begin{align}
 \bar{\bd{B}}^{(1)}(s)\ =&\
[\,\bd{Q}\,,\bar{\bd{\lambda}}^{(1)}(s)\,] 
+ [\,\bd{L}_H(s)\,,\bar{\bd{\lambda}}^{(1)}(s)\,]^{\bar{1}}
+ s\,[\,\bd{L}_H(s)\,,\bar{\bd{\lambda}}^{(1)}(s)\,]^{\bar{2}}
\nonumber\\
=&\
\mathcal{X}\left(
[\,\bd{Q}\,,\bd{\mu}^{(1)}(s)\,] 
+ [\![\,\bd{b}_H(s)\,,\bd{\mu}^{(1)}(s)\,]\!]^{\bar{1}}
+ s\,[\![\,\bd{b}_H(s)\,,\bd{\mu}^{(1)}(s)\,]\!]^{\bar{2}}\right)\
\equiv \mathcal{X}\bar{\bd{b}}^{(1)}(s)\,,
\end{align}
where the double square brackets $[\![\bd{l},\bd{l}']\!]^{\bar{1}}$ and
$[\![\bd{l},\bd{l}']\!]^{\bar{2}}$ are defined by
\begin{align}
 [\![\bd{l},\bd{l}']\!]^{\bar{1}}\ =&\
\sum_{\bar{r},\bar{s}=0}^\infty \Big(\bd{l}|^{2\bar{r}}\,\mathcal{X}\bd{l}'|^{2\bar{s}}
 -(-1)^{|l||l'|}\bd{l}'|^{2\bar{s}}\,\mathcal{X}\bd{l}|^{2\bar{r}}\Big)|^{2\bar{r}+2\bar{s}}\,,\\
 [\![\bd{l},\bd{l}']\!]^{\bar{2}}\ =&\
\sum_{\bar{r},\bar{s}=0}^\infty \Big(\bd{l}|^{2\bar{r}}\mathcal{X}\bd{l}'|^{2\bar{s}}
-(-1)^{|l||l'|}\bd{l}'|^{2\bar{s}}\mathcal{X}\bd{l}|^{2\bar{r}}\Big)|^{2\bar{r}+2\bar{s}-2}\,.
\end{align}
Similarly solving the higher order products recursively, we obtain
\begin{subequations} 
 \begin{align}
\bd{\mu}^{(\bar{n}+1)}(s)\ =&\ \bar{\xi}\circ\left(\partial_s\bar{\bd{b}}^{(\bar{n})}(s)
+ \sum_{\bar{n}'=0}^{\bar{n}-1}[\![\,\bar{\bd{b}}^{(\bar{n}-\bar{n}')}(s)\,,
\bd{\mu}^{(\bar{n}'+1)}(s)\,]\!]^{\bar{2}}\right)\,,
\label{bar mu}\\
(\bar{n}+1)\bar{\bd{b}}^{(\bar{n}+1)}(s)\ =&\ [\,\bd{Q}\,,\bd{\mu}^{(\bar{n}+1)}(s)\,]
\nonumber\\
&\
+ \sum_{\bar{n}'=0}^{\bar{n}}[\![\,\bar{\bd{b}}^{(\bar{n}-\bar{n}')}(s)\,,
\bd{\mu}^{(\bar{n}'+1)}(s)\,]\!]^{\bar{1}}
+ \sum_{\bar{n}'=0}^{\bar{n}}s\,[\![\,\bar{\bd{b}}^{(\bar{n}-\bar{n}')}(s)\,,
\bd{\mu}^{(\bar{n}'+1)}(s)\,]\!]^{\bar{2}}\,,
\label{bar b}
\end{align}
\end{subequations}
where
\begin{equation}
 \bar{\bd{\lambda}}^{(\bar{n}+1)}(s)\ =\ \mathcal{X}\bd{\mu}^{(\bar{n}+1)}(s)\,,\qquad
\bar{\bd{B}}^{(\bar{n}+1)}(s)\ =\ \mathcal{X}\bar{\bd{b}}^{(\bar{n}+1)}(s)\,.
\label{bar lambda}
\end{equation}
Note that the factor $\mathcal{X}$ for the left-moving sector can still be pulled 
out after repeating the procedure for the right-moving sector.
The initial condition $\bar{\bd{B}}^{(0)}(s)=\bd{L}^{(0)}_H(s)$ fixes the structure
of the left-moving picture number as
\begin{subequations} 
 \begin{align}
\bd{\mu}^{(\bar{n}+1)}_{\bar{m}+\bar{n}+\bar{r}+2}\big{|}^{2\bar{r}}\ 
=&\ \sum_{n,r=0}^\infty \delta_{n+r,\bar{m}+\bar{n}+\bar{r}+1}
\bd{\mu}^{(n,\bar{n}+1)}_{\bar{m}+\bar{n}+\bar{r}+2}\big{|}^{(2r,2\bar{r})}\,,
\label{hidden left mu}
\\
\bar{\bd{b}}^{(\bar{n})}_{\bar{m}+\bar{n}+\bar{r}+1}\big{|}^{2\bar{r}}\
=&\ \sum_{n,r=0}^\infty \delta_{n+r,\bar{m}+\bar{n}+\bar{r}}
\bar{\bd{b}}^{(n,\bar{n})}_{\bar{m}+\bar{n}+\bar{r}+1}\big{|}^{(2r,2\bar{r})}\,.
\label{hidden left}
\end{align}
\end{subequations}
All the $\bar{\bd{B}}(s,t)$ and $\bar{\bd{\lambda}}(s,t)$ are in
$\mathcal{H}_{\bar{m}}$\,, satisfying
$[\bd{\eta},\bar{\bd{B}}(s,t)]=[\bd{\eta},\bar{\bd{\lambda}}(s,t)]=0$\,,
 and now cyclic with respect to the symplectic form $\omega_{\bar{m}}$ 
by construction.
We have obtained a cyclic $L_\infty$ algebra 
$(\mathcal{H}_{\bar{m}},\omega_{\bar{m}},\bar{\bd{D}}-\bar{\bd{C})}$\,.
Finally, the cohomomorphism 
\begin{equation}
\pi_1 \hat{\bar{\bd{F}}}^{-1}\ 
=\ \pi_1\idSH - \bar{\Xi}\pi_1^{(*,1)}\bar{\bd{B}}\,,
\end{equation}
generates the similarity transformation which provides 
an $L_\infty$ algebra $\bd{L}$ in the small Hilbert space,
satisfying $[\bd{\eta},\bd{L}]=[\bar{\bd{\eta}},\bd{L}]=0$\,:
\begin{align}
 \pi_1\bd{L}\ \equiv\ 
\pi_1\hat{\bar{\bd{F}}}^{-1}\bar{\bd{D}}\hat{\bar{\bd{F}}}\
=&\ \pi_1\bd{Q} + \bar{\mathcal{X}}\bar{\bd{B}}\hat{\bar{\bd{F}}}
\nonumber\\
=&\ 
\pi_1\bd{Q}+\mathcal{G}\bd{b}\,,
\end{align}
where
$\bd{b}=\bar{\bd{b}}\hat{\bar{\bd{F}}}$
 with
$\bar{\bd{b}}=\sum_{\bar{p},\bar{r}=0}^\infty 
\bar{\bd{b}}^{(\bar{p})}_{\bar{p}+\bar{r}+1}|^{2\bar{r}}$\,.
We can again show that $\bd{b}$ is cyclic with respect to $\omega_l$
in a similar way to Ref.~\cite{Kunitomo:2019glq}, and thus 
the cyclic $L_\infty$ algebra $(\mathcal{H},\Omega,\bd{L})$ is obtained.

\sectiono{Four-point amplitudes}\label{Four point}

In this section we concretely calculate three typical on-shell physical
amplitudes with four external strings 
in a similar way to Refs.~\cite{Erler:2013xta,Kunitomo:2019glq} to demonstrate
how the type II string field theory we have constructed reproduces 
the first-quantized amplitudes. We take the Siegel gauge defined by the conditions
\begin{equation}
\bd{\psi}_{\NSNS}\ =\ \bd{\psi}_{\RNS}\ =\ \bd{\psi}_{\NSR}\ =\ \bd{\psi}_{\RR}\ =\ 0\,.
\end{equation}
From the action (\ref{action}) we can find that
the propagators in this gauge are given by
\begin{equation}
 \begin{alignedat}{2}
 \Pi_{\NSNS}\ =&\ 
-\frac{b_0^+b_0^-}{L_0^+}\delta(L_0^-)\,,\qquad&
 \Pi_{\RNS}\ =&\ -\frac{b_0^+b_0^-X}{L_0^+}\delta(L_0^-)\,,\\
 \Pi_{\NSR}\ =&\ -\frac{b_0^+b_0^-\bar{X}}{L_0^+}\delta(L_0^-)\,,\qquad&
\Pi_{\RR}\ =&\ -\frac{b_0^+b_0^-X\bar{X}}{L_0^+}\delta(L_0^-)\,,
\end{alignedat} 
\end{equation}
%
which agree with those appearing in the first-quantized
formulation \cite{Witten:2012bh}.

\subsection{Four-(R-R) amplitude}

Let us first consider the case that all the external strings are in 
the R-R sector. 
The first-quantized amplitude is given in the form
\begin{equation}
 \mathcal{A}_4^{1st}(\Phi_1\,,\Phi_2\,,\Phi_3\,,\Phi_4)\
=\ \int d^2z \langle\Phi_1(0)
(b_{-1}^+b_{-1}^-\Phi_2(z))\Phi_3(1)\Phi_4(\infty)\rangle\,,
\end{equation}
where $\Phi_1,\cdots,\Phi_4$
are on-shell physical R-R vertex operators, satisfying $Q\Phi=0$\,, 
in a $(-1/2,-1/2)$ picture.
The correlator is evaluated in the small Hilbert space
on the complex $z$-plane. It is not necessary to add any picture-changing operators
at all. Owing to the same moduli structure as the bosonic closed string,
we can express this using the bosonic closed string products $L^B_n$ as
\begin{align}
 \mathcal{A}_4^{1st}(\Phi_1\,,\Phi_2\,,\Phi_3\,,\Phi_4)\ =&\ \omega_s\Bigg(
\Phi_1\,,\bigg(L^B_3(\Phi_2\,,\Phi_3\,,\Phi_4)
- L^B_2\Big(\Phi_2\,, \frac{b_0^+}{L_0^+}L^B_2(\Phi_3\,,\Phi_4)\Big)
\nonumber\\
&\
- L^B_2\Big(\Phi_3\,, \frac{b_0^+}{L_0^+}L^B_2(\Phi_4\,,\Phi_2)\Big)
- L^B_2\Big(\Phi_4\,, \frac{b_0^+}{L_0^+}L^B_2(\Phi_2\,,\Phi_3)\Big)
\bigg)\Bigg)\,.
\label{4RR 1st}
\end{align}
It should be noted that the moduli integral $b_0^-\delta(L_0^-)
=\int^{2\pi}_0\frac{d\theta}{2\pi}b_0^-\exp(i\theta L_0^-)$ is hidden behind the definition
of the string product.
This amplitude can be regarded as a multi-linear map:
\begin{equation}
 \langle\mathcal{A}_4|\,:\,\mathcal{H}_Q^{\RR}\otimes(\mathcal{H}_Q^{\RR})^{\wedge 3}\,
\longrightarrow\,\mathbb{C}\,,
\end{equation}
where $\mathcal{H}_Q\subset\mathcal{H}$ is the subspace of 
states annihilated by $Q$\,. Putting the string fields $\Phi_1,\cdots,\Phi_4$ out,
we can express Eq.~(\ref{4RR 1st}) as
\begin{equation}
 \langle\mathcal{A}_4^{1st}|\ =\
\langle\omega_s|\,\id\otimes\Bigg(
L^B_3 - L^B_2\bigg(
\id\wedge\frac{b_0^+}{L_0^+}L^B_2\bigg)\Bigg)\,,
\label{4RR map}
\end{equation}
by introducing the bilinear map representation $\langle\omega_s|$ 
of the symplectic form $\omega_s$ defined by
\begin{equation}
 \begin{alignedat}{6}
\mathcal{H}&\otimes\mathcal{H}\qquad & &\stackrel{\langle\omega_s|}{\longrightarrow}&\qquad &\mathbb{C}
\\
&\ \rotatebox{90}{$\in$} & & &  &\rotatebox{90}{$\in$}
\\
 |\Phi_1\rangle&\otimes|\Phi_2\rangle\qquad & &\longmapsto&\qquad \omega_s(\Phi&_1\,,\Phi_2)\,.
\end{alignedat}
\end{equation} 
The expression in Eq.~(\ref{4RR map}) can also be written by
using the coderivations as
\begin{equation}
 \langle\mathcal{A}_4^{1st}|\ =\ 
\langle\omega_s|\,\id\otimes\pi_1^{(1,1)}\Big(
\bd{L}^B_3\big{|}^{(4,4)}_{(2,2)} 
- \bd{L}^B_2\big{|}^{(2,2)}_{(0,0)}\, \frac{b_0^+}{L_0^+}
\bd{L}^B_2\big{|}^{(2,2)}_{(2,2)}\Big)\,.
\label{RR 1st}
\end{equation}
Here, $\frac{b_0^+}{L_0^+}\bd{L}^B_2$ is the coderivation derived from
$\frac{b_0^+}{L_0^+}L^B_2$\,.

From the type II superstring field theory, on the other hand,
the four-(R-R) amplitude is calculated as
\begin{align}
 \mathcal{A}_4(\Phi_1\,,\Phi_2\,,\Phi_3\,,\Phi_4)\ =&\
\omega_s\Bigg(\Phi_1\,,\bigg(
b_3^{(0,0)}(\Phi_2\,,\Phi_3\,,\Phi_4)
- b_2^{(0,0)}\Big(\Phi_2\,,\Pi_{\NSNS}\,c_0^- b_2^{(0,0)}(\Phi_3\,,\Phi_4)\Big)
\nonumber\\
&\
- b_2^{(0,0)}\Big(\Phi_3\,,\Pi_{\NSNS}\,c_0^-
 b_2^{(0,0)}(\Phi_4\,,\Phi_2)\Big)
- b_2^{(0,0)}\Big(\Phi_4\,,\Pi_{\NSNS}\,c_0^- b_2^{(0,0)}(\Phi_2\,,\Phi_3)\Big)
\bigg)\Bigg)
\nonumber\\
=&\
\omega_s\Bigg(\Phi_1\,,\bigg(
b_3^{(0,0)}(\Phi_2\,,\Phi_3\,,\Phi_4)
- b_2^{(0,0)}\Big(\Phi_2\,,\frac{b_0^+}{L_0^+}b_2^{(0,0)}(\Phi_3\,,\Phi_4)\Big)
\nonumber\\
&\
- b_2^{(0,0)}\Big(\Phi_3\,,\frac{b_0^+}{L_0^+}b_2^{(0,0)}(\Phi_4\,,\Phi_2)\Big)
- b_2^{(0,0)}\Big(\Phi_4\,,\frac{b_0^+}{L_0^+}b_2^{(0,0)}(\Phi_2\,,\Phi_3)\Big)
\bigg)\Bigg)\,.
\end{align}
The second equality holds owing to the fact that the string field
$b_2^{(0,0)}(\Phi_1,\Phi_2)$ satisfies
the closed string constraints in Eq.~(\ref{restrict closed}). 
Rewriting by using the coderivations, we find
\begin{equation}
 \langle\mathcal{A}_4|\ =\ \langle\omega_s|\id\otimes\pi_1^{(1,1)}\Big(
\bd{b}_3^{(0,0)}\big{|}^{(4,4)}_{(2,2)} - \bd{b}_2^{(0,0)}\big{|}^{(2,2)}_{(0,0)}\,
\frac{b_0^+}{L_0^+}\bd{b}_2^{(0,0)}\big{|}^{(2,2)}_{(2,2)}\Big)\,. 
\label{4 RR sft amplitude}
\end{equation}
Here, the string products without picture number 
$\bd{b}_n^{(0,0)}$ are equal to the bosonic string products
$\bd{L}^B_n$ by construction.\footnote{This will be used without 
notice hereafter.}
Hence, the string field theory amplitude in Eq.~(\ref{4 RR sft amplitude}) 
certainly agrees with the first-quantized amplitude in Eq.~(\ref{RR 1st}).

\subsection{Two-(NS-R)-two-(R-R) amplitude}

Next, we consider the amplitude with two NS-R strings and two R-R strings, 
which is given in the first-quantized formulation by
\begin{equation}
 \langle\mathcal{A}_4|\
=\ \langle\omega_s|\,X_0\otimes\pi_1^{(0,1)}\Big(
\bd{L}_3^{B}\big{|}^{(2,4)}_{(2,2)} 
- \bd{L}_2^{B}\big{|}^{(0,2)}_{(0,0)}\,
  \frac{b_0^+}{L_0^+}\bd{L}_2^{B}\big{|}^{(2,2)}_{(2,2)}
- \bd{L}_2^{B}\big{|}^{(2,2)}_{(2,0)}\,
  \frac{b_0^+}{L_0^+}\bd{L}_2^{B}\big{|}^{(2,2)}_{(0,2)}
\Big)\,,\label{1st 2 NSR 2 RR}
\end{equation}
in a similar representation to the four-(R-R) amplitude.
Here, $\langle\mathcal{A}_4|$ is a multi-linear map
\begin{equation}
 \langle\mathcal{A}_4|\,:\,\mathcal{H}_Q^{\NSR}\otimes\left(\mathcal{H}_Q^{\NSR}\wedge
(\mathcal{H}_Q^{\RR})^{\wedge 2}\right)\,
\longrightarrow\,\mathbb{C}\,,
\end{equation}
and $X_0=\{Q,\xi\}$\,.
In this case, the amplitude obtained from the type II string field theory is calculated as
\begin{align}
 \langle\mathcal{A}_4|\ 
=&\ \langle\omega_s|\,\id\otimes\pi_1^{{(0,1)}}\left(
\bd{b}_3^{(1,0)}\big{|}^{(2,4)}_{(2,2)} 
- \bd{b}_2^{(1,0)}\big{|}^{(0,2)}_{(0,0)}
  \frac{b_0^+}{L_0^+}\bd{L}_2^B\big{|}^{(2,2)}_{(2,2)}
- \bd{L}_2^B\big{|}^{(2,2)}_{(2,0)}
  \frac{b_0^+X}{L_0^+}\bd{L}_2^B\big{|}^{(2,2)}_{(0,2)}
\right)\,. \label{2 NSR 2 RR}
\end{align}
In our construction given in the previous section, 
the string products without right-moving picture number
$\bd{b}_n^{(1,0)}$ are equal to the heterotic string products 
$(\bd{b}_H)_n^{(1)}=(\bd{B}\hat{\bd{F}})_n^{(1)}$
and are given explicitly by
\begin{subequations} \label{by def 1}
 \begin{align}
\sum_{\bar{r}=0,1} \bd{b}_2^{(1,0)}\big{|}^{(0,2\bar{r})}\ 
=&\  (\bd{B}\hat{\bd{F}})_2^{(1)}\big{|}^0
=\ \bd{B}_2^{(1)}|^0
\nonumber\\
=&\ [\,\bd{Q}\,,\bd{\lambda}_2^{(1)}\big{|}^0]\,,\\
\sum_{\bar{r}=0.1.2} \bd{b}_3^{(1,0)}\big{|}^{(2,2\bar{r})}\ 
=&\ (\bd{B}\hat{\bd{F}})_3^{(1)}\big{|}^2
=\ \ \bd{B}_3^{(1)}\big{|}^2 
+ \bd{B}_2^{(0)}\big{|}^2\,\Xi\pi_1^{(1,*)}\bd{B}_2^{(0)}\big{|}^2
\nonumber\\
=&\ [\,\bd{Q}\,,\bd{\lambda}^{(1)}_3\big{|}^2\,]
+ [\,\bd{B}_2^{(0)}\big{|}^2,\,\bd{\lambda}^{(1)}_2\big{|}^2\,]\big{|}^2
+ \bd{B}_2^{(0)}\big{|}^2\,\Xi\pi_1^{(1,*)}\bd{B}_2^{(0)}\big{|}^2\,,
\end{align}
\end{subequations}
where the last equalities follow from the recursion relation in Eq.~(\ref{B}) with $n=0$ .
If we further note that
\begin{equation}
 \bd{B}_2^{(0)}\big{|}^2\ 
 =\ \sum_{\bar{r}=0,1}\bd{L}_2^B\big{|}^{(2,2\bar{r})}\,,\qquad
\bd{\lambda}_2^{(1)}\big{|}^0\ 
=\ \sum_{\bar{r}=0,1}\bd{\lambda}_2^{(1)}\big{|}^{(0,2\bar{r})}\,,
\qquad
\bd{\lambda}_3^{(1)}\big{|}^2\ 
=\ \sum_{\bar{r}=0,1,2}\bd{\lambda}_3^{(1)}\big{|}^{(2,2\bar{r})}\,,
\end{equation}
the relations in Eq.~(\ref{by def 1}) can be decomposed with respect to 
the Ramond and cyclic Ramond numbers. In particular, we find that
\begin{subequations} 
\begin{align}
\pi_1^{(0,1)}\bd{b}_2^{(1,0)}\big{|}^{(0,2)}_{(0,0)}\ =&\ 
\pi_1^{(0,1)}[\,\bd{Q}\,, \bd{\lambda}^{(1)}_2\big{|}^{(0,2)}_{(0,0)}\,]\,,\\
\pi_1^{(0,1)}\bd{b}_3^{(1,0)}\big{|}^{(2,4)}_{(2,2)}\ =&\ 
\pi_1^{(0,1)}\Big(
[\,\bd{Q}\,,\bd{\lambda}_3^{(1)}\big{|}^{(2,4)}_{(2,2)}\,]
- \bd{\lambda}^{(1)}_2\big{|}^{(0,2)}_{(0,0)}\bd{L}_2^B\big{|}^{(2,2)}_{(2,2)}
+ \bd{L}_2^B\big{|}^{(2,2)}_{(2,0)}\,\Xi\bd{L}_2^B\big{|}^{(2,2)}_{(0,2)}\Big)\,.
\end{align}
\end{subequations}
Substituting this into the string field theory amplitude in Eq.~(\ref{2 NSR 2 RR}) 
and then pulling $\bd{Q}$ out, we can rewrite it as
\begin{align}
 \langle\mathcal{A}_4|\ 
=&\ \langle\omega_l|\,\xi\bar{\xi}\otimes\pi_1^{{(0,1)}}\Big(
[\,\bd{Q}\,,\bd{\lambda}_3^{(1)}\big{|}^{(2,4)}_{(2,2)}\,]
- \bd{\lambda}^{(1)}_2\big{|}^{(0,2)}_{(0,0)}\bd{L}_2^B\big{|}^{(2,2)}_{(2,2)}
+ \bd{L}_2^B\big{|}^{(2,2)}_{(2,0)}\,\Xi\bd{L}_2^B\big{|}^{(2,2)}_{(0,2)}
\nonumber\\
&\hspace{30mm}
- [\,\bd{Q}\,,\bd{\lambda}_2^{(1)}\big{|}^{(0,2)}_{(0,0)}\,]
\frac{b_0^+}{L_0^+}\bd{L}_2^B\big{|}^{(2,2)}_{(2,2)}
- \bd{L}_2^B\big{|}^{(2,2)}_{(2,0)}\frac{b_0^+\{\bd{Q}\,,\Xi\}}{L_0^+}
\bd{L}_2^B\big{|}^{(2,2)}_{(0,2)}
\Big) 
\nonumber\\
=&\ -\langle\omega_l|(\xi\bar{X}_0-X_0\bar{\xi})
\nonumber\\
&\hspace{20mm}
\otimes
\pi_1^{(0,1)}\Big(
\bd{\lambda}_3^{(1)}\big{|}^{(2,4)}_{(2,2)}
-\bd{\lambda}_2^{(1)}\big{|}^{(0,2)}_{(0,0)}
  \frac{b_0^+}{L_0^+}\bd{L}_2^B\big{|}^{(2,2)}_{(2,2)}
-\bd{L}_2^B\big{|}^{(2,2)}_{(2,0)}
  \frac{b_0^+\Xi}{L_0^+}\bd{L}_2^B\big{|}^{(2,2)}_{(0,2)}
\Big)\,,\label{sft amplitude}
\end{align}
except for the terms vanishing when they hit the states in $\mathcal{H}_Q$\,.
Inserting $1=[\eta,\xi]$ or $1=[\bar{\eta},\bar{\xi}]$\,, 
we can find that the amplitude in Eq.~(\ref{sft amplitude}) agrees with the first-quantized one:
\begin{align}
 \langle\mathcal{A}_4|\ 
=&\ \langle\omega_s|X_0\otimes\pi_1^{(0,1)}\Big(
[\,\bd{\eta}\,,\bd{\lambda}_3^{(1)}\big{|}^{(2,4)}_{(2,2)}\,]
- [\,\bd{\eta}\,,\bd{\lambda}_2^{(1)}\big{|}^{(0,2)}_{(0,0)}\,]
\frac{b_0^+}{L_0^+}\bd{L}_2^B\big{|}^{(2,2)}_{(2,2)}
-\bd{L}_2^B\big{|}^{(2,2)}_{(2,0)}\frac{b_0^+}{L_0^+}\bd{L}_2^B\big{|}^{(2,2)}_{(0,2)}
\Big)
\nonumber\\
=&\ \langle\omega_s|X_0\otimes\pi_1^{(0,1)}\Big(
\bd{L}_3^B\big{|}^{(2,4)}_{(2,2)}
- \bd{L}_2^B\big{|}^{(0,2)}_{(0,0)}
\frac{b_0^+}{L_0^+}\bd{L}_2^B\big{|}^{(2,2)}_{(2,2)}
-\bd{L}_2^B\big{|}^{(2,2)}_{(2,0)}
\frac{b_0^+}{L_0^+}\bd{L}_2^B\big{|}^{(2,2)}_{(0,2)}
\Big)\,.
\label{sft ex 2}
\end{align}
In the second equality
we used $[\bar{\bd{\eta}},\bd{\lambda}_n^{(1)}]=[\bar{\bd{\eta}},\bd{L}_n^B]=0$ 
and the recursion relation in Eq.~(\ref{lambda}) with $n=0$\,,
\begin{equation}
[\,\bd{\eta}\,,\bd{\lambda}_3^{(1)}\big{|}^{(2,4)}_{(2,2)}\,]\ =\
 \bd{L}_3^B\big{|}^{(2,4)}_{(2,2)}\,,\qquad
[\,\bd{\eta}\,,\bd{\lambda}_2^{(1)}\big{|}^{(0,2)}_{(0,0)}\,]\ =\
 \bd{L}_2^B\big{|}^{(0,2)}_{(0,0)}\,.
\label{sft ex2}
\end{equation}

\subsection{(NS-NS)-(R-NS)-(NS-R)-(R-R) amplitude}

Finally, let us consider the case with four external strings coming
from the four different sectors.
The first-quantized amplitude is given by
\begin{align}
 \langle\mathcal{A}_4|\ =&\ 
\langle\omega_s|\,X_0\bar{X}_0\otimes\pi_1^{(0,0)}\Big(
\bd{L}_3^{B}\big{|}^{(2,2)}_{(2,2)} 
- \bd{L}_2^B\big{|}^{(2,0)}_{(2,0)}\,
\frac{b_0^+}{L_0^+}\bd{L}_2^B\big{|}^{(2,2)}_{(0,2)}
\nonumber\\
&\hspace{40mm}
- \bd{L}_2^B\big{|}^{(0,2)}_{(0,2)}\,
\frac{b_0^+}{L_0^+}\bd{L}_2^B\big{|}^{(2,2)}_{(2,0)}
- \bd{L}_2^B\big{|}^{(2,2)}_{(2,2)}\,
\frac{b_0^+}{L_0^+}\bd{L}_2^B\big{|}^{(2,2)}_{(0,0)}
\Big)P
\,;\label{1st NSNS RNS NSR RR}
\end{align}
as a multi-linear map,
\begin{equation}
 \langle\mathcal{A}_4|\,:\,\mathcal{H}_Q^{\NSNS}\otimes\left(
\mathcal{H}_Q^{\RNS}\wedge
\mathcal{H}_Q^{\NSR}\wedge
\mathcal{H}_Q^{\RR}
\right)\,
\longrightarrow\,\mathbb{C}\,.
\end{equation}
Here, $P$ is the projection operator onto 
$\mathcal{H}_Q^{\RNS}\wedge\mathcal{H}_Q^{\NSR}\wedge\mathcal{H}_Q^{\RR}$
necessary to distinguish it from 
$\mathcal{H}_Q^{\NSNS}\wedge\mathcal{H}_Q^{\RR}\wedge\mathcal{H}_Q^{\RR}$\,,
both of which contain two left-moving Ramond states and two right-moving Ramond states.
The string field theory amplitude in this case
is calculated as
\begin{align}
 \langle\mathcal{A}_4|\ 
=&\ \langle\omega_s|\,\id\otimes\pi_1^{{(0,0)}}\Big(
\bd{b}_3^{(1,1)}\big{|}^{(2,2)}_{(2,2)} 
- \bd{b}_2^{(0,1)}\big{|}^{(2,0)}_{(2,0)}
\frac{b_0^+X}{L_0^+}\bd{L}_2^B\big{|}^{(2,2)}_{(0,2)}
\nonumber\\
&\hspace{30mm} 
- \bd{b}_2^{(1,0)}\big{|}^{(0,2)}_{(0,2)}
\frac{b_0^+\bar{X}}{L_0^+}\bd{L}_2^B\big{|}^{(2,2)}_{(2,0)}
- \bd{L}_2^B\big{|}^{(2,2)}_{(2,2)}
\frac{b_0^+X\bar{X}}{L_0^+}\bd{L}_2^B\big{|}^{(2,2)}_{(0,0)}
\Big)P\,. \label{NSNS RNS NSR RR}
\end{align}
Using $\bd{b}=\bar{\bd{b}}\hat{\bar{\bd{F}}}$ and
the recursion relation in Eq.~(\ref{bar b}), we have
\begin{subequations}  
\begin{align}
 \sum_{p=0,1}\bd{b}_2^{(p,1)}\big{|}^{(2(1-p),0)}\ 
=&\ (\bar{\bd{b}}\hat{\bar{\bd{F}}})_2^{(1)}\big{|}^0
=\ \bar{\bd{b}}_2^{(1)}\big{|}^0 
\nonumber\\ 
=&\ [\,\bd{Q}\,,\bd{\mu}_2^{(1)}\big{|}^0\,]\,,\\
 \sum_{p=0,1,2}\bd{b}_3^{(p,1)}\big{|}^{(2(2-p),2)}\ 
=&\ (\bar{\bd{b}}\hat{\bar{\bd{F}}})_3^{(1)}\big{|}^2
=\ \bar{\bd{b}}_3^{(1)}\big{|}^2 
+ \bar{\bd{b}}_2^{(0)}\big{|}^2\,\bar{\Xi}\pi_1^{(*,1)}
\mathcal{X}\bar{\bd{b}}_2^{(0)}\big{|}^2
\nonumber\\
=&\ [\,\bd{Q}\,,\bd{\mu}_3^{(1)}\big{|}^2\,] + 
[\![\,\bar{\bd{b}}_2^{(0)}\big{|}^2,\,\bd{\mu}_2^{(1)}\big{|}^0\,]\!]^{\bar{1}}
+ \bar{\bd{b}}_2^{(0)}\big{|}^2\,\bar{\Xi}\pi_1^{(*,1)}
\mathcal{X}\bar{\bd{b}}_2^{(0)}\big{|}^2\,,
\end{align}
\end{subequations}
from which we can find that
\begin{subequations} 
 \begin{align}
\pi_1^{(0,0)}\bd{b}_2^{(0,1)}\big{|}^{(2,0)}_{(2,0)}\ =&\ 
\pi_1^{(0,0)}[\,\bd{Q},\bd{\mu}_2^{(0,1)}\big{|}^{(2,0)}_{(2,0)}\,]\,,\\
 \pi_1^{(0,0)}\bd{b}_3^{(1,1)}\big{|}^{(2,2)}_{(2,2)}P\ =&\
\pi_1^{(0,0)}\Big([\,\bd{Q},\bd{\mu}_3^{(1,1)}\big{|}^{(2,2)}_{(2,2)}\,]
- \bd{\mu}_2^{(0,1)}\big{|}^{(2,0)}_{(2,0)}X\bd{L}_2^B\big{|}^{(2,2)}_{(0,2)}
\nonumber\\
&\hspace{13mm}
+\bd{L}_2^B\big{|}^{(2,2)}_{(2,2)}X\bar{\Xi}\,\bd{L}_2^B\big{|}^{(2,2)}_{(0,0)}
+\bar{\bd{b}}_2^{(1,0)}\big{|}^{(0,2)}_{(0,2)}\,\bar{\Xi}\,\bd{L}_2^B\big{|}^{(2,2)}_{(2,0)}
\Big)P
\end{align}
\end{subequations}
by decomposing it with respect to the cyclic Ramond and Ramond numbers.
Then, the string field theory amplitude 
in Eq.~(\ref{NSNS RNS NSR RR}) can be rewritten as
\begin{align} 
\langle\mathcal{A}_4|\ 
=&\
\langle\omega_l|\xi\bar{\xi}\otimes\pi_1^{(0,0)}\Big(
\pi_1^{(0,0)}\Big([\,\bd{Q},\bd{\mu}_3^{(1,1)}\big{|}^{(2,2)}_{(2,2)}\,]
- \bd{\mu}_2^{(0,1)}\big{|}^{(2,0)}_{(2,0)}X\bd{L}_2^B\big{|}^{(2,2)}_{(0,2)}
\nonumber\\
&\
+\bd{L}_2^B\big{|}^{(2,2)}_{(2,2)}X\bar{\Xi}\,\bd{L}_2^B\big{|}^{(2,2)}_{(0,0)}
+\bar{\bd{b}}_2^{(1,0)}\big{|}^{(0,2)}_{(0,2)}\bar{\Xi}\,\bd{L}_2^B\big{|}^{(2,2)}_{(2,0)}
- [\,\bd{Q},\bd{\mu}_2^{(0,1)}\big{|}^{(2,0)}_{(2,0)}\,]
\frac{b_0^+X}{L_0^+}\bd{L}_2^B\big{|}^{(2,2)}_{(0,2)}
\nonumber\\
&\
- \bd{b}_2^{(1,0)}\big{|}^{(0,2)}_{(0,2)}
\frac{b_0^+\{\bd{Q}\,,\bar{\Xi}\}}{L_0^+}\bd{L}_2^B\big{|}^{(2,2)}_{(2,0)}
- \bd{L}_2^B\big{|}^{(2,2)}_{(2,2)}
\frac{b_0^+X\{\bd{Q}\,,\bar{\Xi}\}}{L_0^+}\bd{L}_2^B\big{|}^{(2,2)}_{(0,0)}
\Big)P
\nonumber\\
=&\ -\langle\omega_l|(\xi\bar{X}_0-X_0\bar{\xi})\otimes\pi_1^{(0,0)}\Big(
\bd{\mu}_3^{(1,1)}\big{|}^{(2,2)}_{(2,2)}
- \bd{\mu}_2^{(0,1)}\big{|}^{(2,0)}_{(2,0)}
\frac{b_0^+X}{L_0^+}\bd{L}_2^B\big{|}^{(2,2)}_{(0,2)}
\nonumber\\
&\hspace{40mm}
- \bd{b}_2^{(1,0)}\big{|}^{(0,2)}_{(0,2)}
\frac{b_0^+\bar{\Xi}}{L_0^+}\bd{L}_2^B\big{|}^{(2,2)}_{(2,0)}
- \bd{L}_2^B\big{|}^{(2,2)}_{(2,2)}
\frac{b_0^+X\bar{\Xi}}{L_0^+}\bd{L}_2^B\big{|}^{(2,2)}_{(0,0)}
\Big)P\,,
\end{align}
except for the terms which vanish when they hit the states in $\mathcal{H}_Q$\,.
Inserting $1=\{\bar{\eta},\bar{\xi}\}$ or $1=\{\eta,\xi\}$\,, 
we find that
\begin{align}
 \langle\mathcal{A}_4|\ =&\
\langle\omega_s|\,\bar{X}_0\otimes\pi_1^{(0,0)}\Big(
(\bd{b}_H)_3^{(1)}\big{|}^{(2,2)}_{(2,2)}
- \bd{L}_2^B\big{|}^{(2,0)}_{(2,0)}
\frac{b_0^+X}{L_0^+}\bd{L}_2^B\big{|}^{(2,2)}_{(0,2)}
\nonumber\\
&\hspace{40mm}
- (\bd{b}_H)_2^{(1)}\big{|}^{(0,2)}_{(0,2)}
\frac{b_0^+}{L_0^+}\bd{L}_2^B\big{|}^{(2,2)}_{(2,0)}
- \bd{L}_2^B\big{|}^{(2,2)}_{(2,2)}
\frac{b_0^+X}{L_0^+}\bd{L}_2^B\big{|}^{(2,2)}_{(0,0)}
\Big)P
\label{NSNS RNS NSR RR 2}
\end{align}
by using
\begin{equation}
 [\,\bd{\eta},\bd{\mu}\,]\ =\ [\,\bd{\eta},\bd{L}^B\,]\ =\ 0\,,
\end{equation}
and
\begin{subequations}
\begin{align}
 [\,\bar{\bd{\eta}}\,,\bd{\mu}_3^{(1,1)}\big{|}^{(2,2)}_{(2,2)}\,]\ =&\
\bar{\bd{b}}_3^{(1,0)}\big{|}^{(2,2)}_{(2,2)}\
=\ (\bd{b}_H)^{(1)}_3\big{|}^{(2,2)}_{(2,2)}\,,\\
 [\,\bar{\bd{\eta}}\,,\bd{\mu}_2^{(0,1)}\big{|}^{(2,0)}_{(2,0)}\,]\ =&\
\bar{\bd{b}}_2^{(0,0)}\big{|}^{(2,0)}_{(2,0)}\
=\ \bd{L}^B_2\big{|}^{(2,0)}_{(2,0)}\,,
\end{align}
\end{subequations}
following from Eq.~(\ref{lambda bar 1}).

We can repeat a similar procedure for the left-moving sector. 
Using $\bd{b}_H=\bd{B}\hat{\bd{F}}$\,, we have, in particular,
\begin{subequations} 
  \begin{align}
 (\bd{b}_H)_2^{(1)}\big{|}^0\ =&\ (\bd{B}\hat{\bd{F}})_2^{(1)}\big{|}^0\
=\ \bd{B}_2^{(1)}\big{|}^0
\nonumber\\
=&\ [\,\bd{Q}\,,\,\bd{\lambda}_2^{(1)}\big{|}^0\,]\,,\\
  (\bd{b}_H)_3^{(1)}\big{|}^2\ =&\ (\bd{B}\hat{\bd{F}})_3^{(1)}\big{|}^2\
=\ \bd{B}_3^{(1)}\big{|}^2 
+ \bd{B}_2^{(0)}\big{|}^2\,\Xi\pi_1^{(1,*)}\bd{B}_2^{(0)}\big{|}^2
\nonumber\\
=&\ [\,\bd{Q}\,,\,\bd{\lambda}_3^{(1)}\big{|}^2\,]
+ [\,\bd{B}_2^{(0)}\big{|}^2,\,\bd{\lambda}_2^{(1)}\big{|}^0\,]\big{|}^2
+ \bd{B}_2^{(0)}\big{|}^2\,\Xi\pi_1^{(1,*)}\bd{B}_2^{(0)}\big{|}^2\,,
 \end{align}
\end{subequations}
by using Eq.~(\ref{B}). Decomposing these with respect to the cyclic Ramond and
Ramond numbers, we find that
\begin{subequations} 
 \begin{align}
 \pi_1^{(0,0)}(\bd{b}_H)_2^{(1)}\big{|}^{(0,2)}_{(0,2)}\ 
=&\ \pi_1^{(0,0)}[\,\bd{Q}\,,\bd{\lambda}_2^{(1)}\big{|}^{(0,2)}_{(0,2)}\,]\,,\\
\pi_1^{(0,0)} (\bd{b}_H)_3^{(1)}\big{|}^{(2,2)}_{(2,2)}P\ 
=&\ \pi_1^{(0,0)}\Big(
[\,\bd{Q}\,,\bd{\lambda}_3^{(1)}\big{|}^{(2,2)}_{(2,2)}\,]
-\bd{\lambda}_2^{(1)}\big{|}^{(0,2)}_{(0,2)}\,\bd{L}_2^B\big{|}^{(2,2)}_{(2,0)}
\nonumber\\
&\hspace{15mm} 
+ \bd{L}_2^B\big{|}^{(2,0)}_{(2,0)}\,\Xi\bd{L}_2^B\big{|}^{(2,2)}_{(0,2)}
+ \bd{L}_2^B\big{|}^{(2,2)}_{(2,2)}\,\Xi\bd{L}_2^B\big{|}^{(2,2)}_{(0,0)}\Big)P\,.
\end{align}
\end{subequations}
Thanks to these relations, the amplitude 
in Eq.~(\ref{NSNS RNS NSR RR 2}) can be further rewritten as
\begin{align}
 \langle\mathcal{A}_4|\ =&\
\langle\omega_l|\,\bar{X}_0\xi\bar{\xi}\otimes\pi_1^{(0,0)}\Big(
[\,\bd{Q}\,,\bd{\lambda}_3^{(1)}\big{|}^{(2,2)}_{(2,2)}\,]
-\bd{\lambda}_2^{(1)}\big{|}^{(0,2)}_{(0,2)}\,\bd{L}_2^B\big{|}^{(2,2)}_{(2,0)}
\nonumber\\
&\hspace{35mm}
+ \bd{L}_2^B\big{|}^{(2,0)}_{(2,0)}\,\Xi\bd{L}_2^B\big{|}^{(2,2)}_{(0,2)}
+ \bd{L}_2^B\big{|}^{(2,2)}_{(2,2)}\,\Xi\bd{L}_2^B\big{|}^{(2,2)}_{(0,0)}
\nonumber\\
&\hspace{35mm}
- \bd{L}_2^B\big{|}^{(2,0)}_{(2,0)}
\frac{b_0^+\{Q\,,\Xi\}}{L_0^+}\bd{L}_2^B\big{|}^{(2,2)}_{(0,2)}
- [\,\bd{Q}\,,\bd{\lambda}_2^{(1)}\big{|}^{(0,2)}_{(0,2)}\,]
\frac{b_0^+}{L_0^+}\bd{L}_2^B\big{|}^{(2,2)}_{(2,0)}
\nonumber\\
&\hspace{35mm}
- \bd{L}_2^B\big{|}^{(2,2)}_{(2,2)}
\frac{b_0^+\{Q\,,\Xi\}}{L_0^+}\bd{L}_2^B\big{|}^{(2,2)}_{(0,0)}
\Big)P
\nonumber\\
=&\
- \langle\omega_l|\,\bar{X}_0(\xi\bar{X}_0-X_0\bar{\xi})\otimes\pi_1^{(0,0)}\Big(
\bd{\lambda}_3^{(1)}\big{|}^{(2,2)}_{(2,2)}
- \bd{L}_2^B\big{|}^{(2,0)}_{(2,0)}
\frac{b_0^+\Xi}{L_0^+}\bd{L}_2^B\big{|}^{(2,2)}_{(0,2)}
\nonumber\\
&\hspace{50mm}
- \bd{\lambda}_2^{(1)}\big{|}^{(0,2)}_{(0,2)}
\frac{b_0^+}{L_0^+}\bd{L}_2^B\big{|}^{(2,2)}_{(2,0)}
- \bd{L}_2^B\big{|}^{(2,2)}_{(2,2)}
\frac{b_0^+\Xi}{L_0^+}\bd{L}_2^B\big{|}^{(2,2)}_{(0,0)}
\Big)P\,,
\end{align}
except for the terms vanishing when they hit the states in $\mathcal{H}_Q$\,.
Again, inserting $1=\{\eta,\xi\}=\{\bar{\eta},\bar{\xi}\}$\,,
the string field theory amplitude eventually becomes
\begin{align}
 \langle\mathcal{A}_4|\ =&\
\langle\omega_s|\,X_0\bar{X}_0\otimes\pi_1^{(0,0)}\Big(
\bd{L}_3^B\big{|}^{(2,2)}_{(2,2)}
- \bd{L}_2^B\big{|}^{(2,0)}_{(2,0)}
\frac{b_0^+}{L_0^+}\bd{L}_2^B\big{|}^{(2,2)}_{(0,2)}
\nonumber\\
&\hspace{50mm}
- \bd{L}_2^B\big{|}^{(0,2)}_{(0,2)}
\frac{b_0^+}{L_0^+}\bd{L}_2^B\big{|}^{(2,2)}_{(2,0)}
- \bd{L}_2^B\big{|}^{(2,2)}_{(2,2)}
\frac{b_0^+}{L_0^+}\bd{L}_2^B\big{|}^{(2,2)}_{(0,0)}
\Big)P\,,
\end{align}
using $[\bd{\eta},\bd{\lambda}^{(1)}]=\bd{L}^B$\,.
This reproduces the first-quantized amplitude 
of Eq.~(\ref{1st NSNS RNS NSR RR}).

\sectiono{Relation to the WZW-like formulation}\label{relation to WZW}

So far we have constructed a complete gauge-invariant action for 
the type II superstring field theory in the small Hilbert space
based on the cyclic $L_\infty$ structure. 
In open superstring field theory \cite{Erler:2016ybs}
and heterotic string field theory \cite{Kunitomo:2019glq}, we can map it to 
a gauge-invariant action in the WZW-like formulation through a field 
redefinition. In this section we consider whether it is also
possible to construct a complete
WZW-like action in a similar way for the type II superstring field theory.

Here, let us consider the restriction of the construction to the pure NS-NS sector.
If we define generating functions by
\begin{subequations}  
\begin{align}
\bd{L}_H(s,t)\big{|}^{(0,0)}\ \equiv\ \big(\bd{Q}+\bd{B}(s,t)\big)\big{|}^{(0,0)}\ =&\ 
\big(\bd{Q}+\sum_{m,n=0}^\infty s^m t^n \bd{B}^{(n)}_{m+n+1}\big)\big{|}^{(0,0)}\,,\\
\bd{L}(s,t)\big{|}^{(0,0)}\ \equiv\ \big(\bd{Q} + \bar{\bd{B}}(s,t)\big)\big{|}^{(0,0)}\ =&\ 
\big(\bd{Q} + \sum_{\bar{m},\bar{n}=0}^\infty s^{\bar{m}} t^{\bar{n}} 
\bar{\bd{B}}^{(\bar{n})}_{\bar{m}+\bar{n}+1}\big)\big{|}^{(0,0)}\,,
\end{align} 
\end{subequations}
they are the generating functions of $\bd{L}_H|^{(0,0)}$ and $\bd{L}|^{(0,0)}$ 
since the cohomomorphisms $\hat{\bd{F}}$ and $\hat{\bar{\bd{F}}}$ reduce to 
the identity in the NS-NS sector. 
The string products in the NS-NS action 
can be obtained by $\bd{L}|^{(0,0)}=\bd{L}(0,1)|^{(0,0)}$\,. 
These coderivations satisfy the $L_\infty$ relations
\begin{subequations}\label{L infinities} 
 \begin{align}
 [\,\bd{L}_H(s,t)\big{|}^{(0,0)},\,\bd{L}_H(s,t)\big{|}^{(0,0)}\,]\ =&\ 0\,,\\
[\,\bd{L}(s,t)\big{|}^{(0,0)},\,\bd{L}(s,t)\big{|}^{(0,0)}\,]\ =&\ 0\,,
 \end{align}
\end{subequations}
and both of them are closed in the small Hilbert space:
\begin{equation}
 [\,\bd{\eta}\,,\bd{L}_H(s,t)\big{|}^{(0,0)}\,]\ =\ 
 [\,\bar{\bd{\eta}}\,,\bd{L}_H(s,t)\big{|}^{(0,0)}\,]\ =\
 [\,\bd{\eta}\,,\bd{L}(s,t)\big{|}^{(0,0)}\,]\ =\
 [\,\bar{\bd{\eta}}\,,\bd{L}(s,t)\big{|}^{(0,0)}\,]\ =\ 0\,.
\end{equation}
The $L_\infty$ relations in Eq.~(\ref{L infinities}) follow from the differential equations
\begin{subequations}\label{diff for LH} 
\begin{align}
 \partial_t\bd{L}_H(s,t)\big{|}^{(0,0)}\ =&\ 
[\,\bd{L}_H(s,t)\big{|}^{(0,0)}\,,\bd{\lambda}(s,t)\big{|}^{(0,0)}\,]\,,
\label{diff for LH 1}\\
\partial_s\bd{L}_H(s,t)\big{|}^{(0,0)}\ =&\ [\,\bd{\eta}\,,\bd{\lambda}(s,t)\big{|}^{(0,0)}\,]\,,
\label{diff for LH 2}
\end{align}
\end{subequations}
and
\begin{subequations}\label{diff for L}  
\begin{align}
 \partial_t\bd{L}(s,t)\big{|}^{(0,0)}\ =&\
 [\,\bd{L}(s,t)\big{|}^{(0,0)}\,,\bar{\bd{\lambda}}(s,t)\big{|}^{(0,0)}\,]\,,
\label{diff for L 1}\\
\partial_s\bd{L}(s,t)\big{|}^{(0,0)}\ =&\ 
[\,\bar{\bd{\eta}}\,, \bar{\bd{\lambda}}(s,t)\big{|}^{(0,0)}\,]\,,
\label{diff for L 2}
\end{align}
\end{subequations}
derived from Eqs.~(\ref{diff left}) and (\ref{diff right}), respectively,
where $\bd{\lambda}(s,t)|^{(0,0)}$ and $\bar{\bd{\lambda}}(s,t)|^{(0,0)}$
are the similar restrictions of the gauge products in Eq.~(\ref{gf lambda})
and (\ref{gf lambda bar}) to the NS-NS sector.
The required string products are obtained in two steps:
first, we recursively solve Eq.~(\ref{diff for LH}) starting from the initial condition 
$\bd{L}_H(s,0)|^{(0,0)}=\bd{L}_B(s)|^{(0,0)}$, and then solve Eq.~(\ref{diff for L})
with the initial condition $\bd{L}(s,0)|^{(0,0)}=\bd{L}_H(0,s)|^{(0,0)}$\,.
This is nothing but the asymmetric construction proposed in Ref.~\cite{Erler:2014eba}, and thus
the string and gauge products we constructed reduce in the NS-NS sector
to those obtained by their asymmetric construction \cite{Erler:2014eba}.

Using the fact that $\bd{L}(s,t)|^{(0,0)}$ satisfies the differential equation 
in Eq.~(\ref{diff for L}),
we can show that the string products restricted in the NS-NS sector 
$\bd{L}|^{(0,0)}=\bd{L}(0,1)|^{(0,0)}$ can be written 
in the form of the similarity transformation \cite{Erler:2016ybs,Kunitomo:2019glq} as
 \begin{equation}
\bd{L}|^{(0,0)}\ 
=\ \bd{Q} + \bar{\bd{B}}(0,1)\big{|}^{(0,0)}\ =\ \hat{\bd{g}}^{-1}\bd{Q}\hat{\bd{g}}\,,
\end{equation}
by the cohomomorphism
\begin{equation}
 \hat{\bd{g}}\ =\ \vec{\mathcal{P}}\exp\left(\int^1_0dt\bar{\bd{\lambda}}(0,t)|^{(0,0)}\right)\,.
\label{cohomomorphism}
\end{equation}

Due to the commutativity $[\bd{\eta},\bar{\bd{\lambda}}(0,t)]=0$  that holds
by construction, however, it transforms $\bd{\eta}$ and $\bar{\bd{\eta}}$ 
asymmetrically.
The constraints $\eta\Phi_{\NSNS}=\bar{\eta}\Phi_{\NSNS}=0$\,, 
restricting $\Phi_{\NSNS}$ 
to the small Hilbert space, are mapped to 
\begin{equation}
 \eta\,\pi_1\hat{\bd{g}}(e^{\wedge\Phi_{\NSNS}})\ =\ 
\pi_1\bd{L}^{\bar{\eta}}(e^{\wedge\pi_1\hat{\bd{g}}(e^{\wedge\Phi_{\NSNS}})}) =\ 0\,,
\end{equation}
with $\bd{L}^{\bar{\bd{\eta}}}=\hat{\bd{g}}\bar{\bd{\eta}}\hat{\bd{g}}^{-1}$\,.
Using this map, therefore, we can only obtain a \textit{half}-WZW-like
formulation, in which the NS-NS string field $V$
has the ghost and picture numbers
$(1,0)$ and $(-1,0)$\,, respectively, and takes value in the medium Hilbert space 
introduced in Eq.~(\ref{symplectic forms}): $V\in\mathcal{H}_{\bar{m}}$\,.
If we identify the string field through the map
\begin{equation}
 \pi_1\hat{\bd{g}}(e^{\wedge\Phi_{\NSNS}})\ =\ G_{\bar{\eta}}(V)\,,
\label{identification NSNS}
\end{equation}
the pure-gauge string field $G_{\bar{\eta}}(V)$ satisfies the asymmetric 
Maurer-Cartan equations
\begin{equation}
 \eta G_{\bar{\eta}}(V)\ =\ 0\,,\qquad
 \bd{L}^{\bar{\eta}}(e^{\wedge G_{\bar{\eta}}(V)})\ =\ 0\,,
\end{equation}
and thus is given by the one used in heterotic string field theory 
\cite{Goto:2016ckh,Kunitomo:2019glq}.\footnote{The role of $\eta$ in 
heterotic string field theory is played here by $\bar{\eta}$\,.}
The string fields in the other sectors are simply identified in two
formulations.
We denote the string fields of the R-NS, NS-R and R-R sectors
in the half-WZW-like formulation as $\Psi$\,, $\bar{\Psi}$ and $\Sigma$\,, 
respectively, to distinguish which formulation they belong to:
\begin{equation}
\Phi_{\RNS}\ =\ \Psi\,,\qquad
\Phi_{\NSR}\ =\ \bar{\Psi}\,,\qquad
\Phi_{\RR}\ =\ \Sigma\,.
\label{identification others}
\end{equation}

The half-WZW-like formulation obtained in this way is the dual (in the sense that
the roles of $Q$ and $\bar{\eta}$ are exchanged) to that given in 
Ref.~\cite{Matsunaga:2014wpa}. It is not 
the completely WZW-like formulation defined using the (whole) large Hilbert space
$\mathcal{H}_l$\,, but we also construct here an action and 
a gauge transformation to complete the story.
First of all, we rewrite the action (\ref{action}) in the WZW-like
form by extending the NS-NS string field $\Phi_{\NSNS}$ to 
$\Phi_{\NSNS}(t)$ with $t\in[0,1]$ satisfying $\Phi_{\NSNS}(0)=0$
and $\Phi_{\NSNS}(1)=\Phi_{\NSNS}$\,. Using the cyclicity, we find
that
\begin{align}
 S\ =&\
\int^1_0 dt\,
\omega_m(\bar{\xi}\partial_t\Phi_{\NSNS}(t)\,, 
\pi_1^{(0,0)}\bd{L}(e^{\wedge\Phi_{\NSNS}(t)}))
\nonumber\\
&\
+ \frac{1}{2}\omega_s(\Phi_{\RNS}\,, YQ\Phi_{\RNS})
+ \frac{1}{2}\omega_s(\Phi_{\NSR}\,, \bar{Y}Q\Phi_{\NSR})
+ \frac{1}{2}\omega_s(\Phi_{\RR}\,, \mathcal{Y}\bar{\mathcal{Y}}Q\Phi_{\RR})
\nonumber
 \end{align}
\begin{align}
&\
+ \sum_{r=0}^\infty\frac{1}{(2r+2)!}\Big(
\omega_s\big(\Phi_{\RNS}\,,\pi_1^{(1,0)}\bd{b}(e^{\wedge\Phi_{\NSNS}}
\wedge{\Phi_{\RNS}}^{\wedge 2r+1})\big)
\nonumber\\
&\hspace{32mm}
+ \omega_s\big(\Phi_{\NSR}\,,\pi_1^{(0,1)}\bd{b}(e^{\wedge\Phi_{\NSNS}}
\wedge{\Phi_{\NSR}}^{\wedge 2r+1})\big)
\nonumber\\
&\
\nonumber\\
&\hspace{37mm}
+ \omega_s\big(\Phi_{\RR}\,,\pi_1^{(1,1)}\bd{b}(e^{\wedge\Phi_{\NSNS}}
\wedge{\Phi_{\RR}}^{\wedge 2r+1})\big)\Big)
\nonumber\\
&\
+\sum_{r_1,r_2=0}^\infty\frac{1}{(2r_1+2)!(2r_2+2)!}
\nonumber\\
&\hspace{20mm}\times
\Bigg[\frac{1}{2}\Big(
\omega_s\big(\Phi_{\RNS}\,,\pi_1^{(1,0)}\bd{b}(e^{\wedge\Phi_{\NSNS}}\wedge
{\Phi_{\RNS}}^{\wedge 2r_1+1}\wedge{\Phi_{\NSR}}^{2r_2+2})\big)
\nonumber\\
&\hspace{30mm} 
+\omega_s\big(\Phi_{\NSR}\,,\pi_1^{(0,1)}\bd{b}(e^{\wedge\Phi_{\NSNS}}\wedge
{\Phi_{\RNS}}^{\wedge 2r_1+2}\wedge{\Phi_{\NSR}}^{2r_2+1})\big)\Big)
\nonumber\\
&\hspace{25mm} 
+\omega_s\big(\Phi_{\RR}\,,\pi_1^{(1,1)}\bd{b}(e^{\wedge\Phi_{\NSNS}}\wedge
{\Phi_{\RNS}}^{\wedge 2r_1+2}\wedge{\Phi_{\RR}}^{\wedge 2r_2+1})\big)
\nonumber\\
&\hspace{25mm} 
+\omega_s\big(\Phi_{\RR}\,,\pi_1^{(1,1)}\bd{b}(e^{\wedge\Phi_{\NSNS}}\wedge
{\Phi_{\NSR}}^{\wedge 2r_1+2}\wedge{\Phi_{\RR}}^{\wedge 2r_2+1})\big)
\Bigg]
\nonumber\\
&\
+ \sum_{r_1,r_2,r_3=0}^\infty
\Bigg[\frac{1}{(2r_1+1)!(2r_2+1)!(2r_3+1)!}
\nonumber\\
&\hspace{10mm}
\times \omega_s\big(\Phi_{\RR}\,,\pi_1^{(1.1)}\bd{b}(e^{\wedge\Phi_{\NSNS}}\wedge
{\Phi_{\RNS}}^{\wedge 2r_1+1}\wedge{\Phi_{\NSR}}^{\wedge 2r_2+1}\wedge
{\Phi_{\RR}}^{\wedge 2r_3})\big)
\nonumber\\
&\hspace{19mm}
+\frac{1}{(2r_1+2)!(2r_2+2)!(2r_3+2)!}
\nonumber\\
&\hspace{10mm}
\times\omega_s\big(\Phi_{\RR}\,,\pi_1^{(1,1)}\bd{b}(e^{\wedge\Phi_{NSNS}}\wedge
{\Phi_{\RNS}}^{\wedge 2r_1+2}\wedge{\Phi_{\NSR}}^{\wedge 2r_2+2}\wedge
{\Phi_{\RR}}^{\wedge 2r_3+1})\big)
\Bigg]\,.
\end{align}
It is mapped to the half-WZW-like action through the identification
in Eqs.(\ref{identification NSNS}) and (\ref{identification others}) as
\begin{align}
 S\ =&\
\int^1_0dt\,\omega_m(B_t(V(t))\,,QG_{\bar{\eta}}(V(t)))
\nonumber\\
&\
+ \frac{1}{2}\omega_s(\Psi\,, YQ\Psi)
+ \frac{1}{2}\omega_s(\bar{\Psi}\,, \bar{Y}Q\bar{\Psi})
+ \frac{1}{2}\omega_s(\Sigma\,, \mathcal{Y}\bar{\mathcal{Y}}Q\Sigma)
\nonumber \\
&\
+ \sum_{r=0}^\infty\frac{1}{(2r+2)!}\Big(
\omega_s\big(\Psi\,,\pi_1^{(1,0)}\tilde{\bd{b}}(e^{\wedge G_{\bar{\eta}}(V)}
\wedge{\Psi}^{\wedge 2r+1})\big)
\nonumber\\
&\hspace{32mm}
+ \omega_s\big(\bar{\Psi}\,,\pi_1^{(0,1)}\tilde{\bd{b}}(e^{\wedge G_{\bar{\eta}}(V)}
\wedge\bar{\Psi}^{\wedge 2r+1})\big)
\nonumber\\
&\hspace{37mm}
+ \omega_s\big(\Sigma\,,\pi_1^{(1,1)}\tilde{\bd{b}}(e^{\wedge G_{\bar{\eta}}(V)}
\wedge \Sigma^{\wedge 2r+1})\big)\Big)
\nonumber
 \end{align}
\begin{align}
&\
+\sum_{r_1,r_2=0}^\infty\frac{1}{(2r_1+2)!(2r_2+2)!}
\nonumber\\
&\hspace{20mm}\times
\Bigg[\frac{1}{2}\Big(
\omega_s\big(\Psi\,,\pi_1^{(1,0)}\tilde{\bd{b}}(e^{\wedge G_{\bar{\eta}}(V)}\wedge
\Psi^{\wedge 2r_1+1}\wedge\bar{\Psi}^{2r_2+2})\big)
\nonumber\\
&\hspace{30mm} 
+\omega_s\big(\bar{\Psi}\,,\pi_1^{(0,1)}\bd{b}(e^{\wedge G_{\bar{\eta}}(V)}\wedge
\Psi^{\wedge 2r_1+2}\wedge\bar{\Psi}^{2r_2+1})\big)\Big)
\nonumber\\
&\hspace{20mm} 
+\omega_s\big(\Sigma\,,\pi_1^{(1,1)}\tilde{\bd{b}}(e^{\wedge G_{\bar{\eta}}(V)}\wedge
\Psi^{\wedge 2r_1+2}\wedge\Sigma^{\wedge 2r_2+1})\big)
\nonumber\\
&\hspace{20mm} 
+\omega_s\big(\Sigma\,,\pi_1^{(1,1)}\tilde{\bd{b}}(e^{\wedge G_{\bar{\eta}}(V)}\wedge
\bar{\Psi}^{\wedge 2r_1+2}\wedge\Sigma^{\wedge 2r_2+1})\big)
\Bigg]
\nonumber\\
&\
+ \sum_{r_1,r_2,r_3=0}^\infty
\Bigg[\frac{1}{(2r_1+1)!(2r_2+1)!(2r_3+1)!}
\nonumber\\
&\
\times \omega_s\big(\Sigma\,,\pi_1^{(1.1)}\tilde{\bd{b}}(e^{\wedge G_{\bar{\eta}}(V)}\wedge
\Psi^{\wedge 2r_1+1}\wedge\bar{\Psi}^{\wedge 2r_2+1}\wedge
\Sigma^{\wedge 2r_3})\big)
\nonumber\\
&\hspace{19mm}
+\frac{1}{(2r_1+2)!(2r_2+2)!(2r_3+2)!}
\nonumber\\
&\
\times\omega_s\big(\Sigma\,,\pi_1^{(1,1)}\tilde{\bd{b}}(e^{\wedge G_{\bar{\eta}}(V)}\wedge
\Psi^{\wedge 2R_1+2}\wedge\bar{\Psi}^{\wedge 2R_2+2}\wedge
\Sigma^{\wedge 2R_3+1})\big)
\Bigg]\,,
\label{WZW-like action}
\end{align}
where $\tilde{\bd{b}}=\hat{\bd{g}}(\bd{b}-\bar{\bd{B}}(0,1)|^{(0,0)})\hat{\bd{g}}^{-1}$\,.
Here, we defined the associated fields as
\begin{align}
 B_d(V(t))\ =\ \pi_1^{(0,0)}\hat{\bd{g}}\bd{\bar{\xi}}_{\bd{d}}
(e^{\wedge\Phi_{\NSNS}(t)})\,,
\label{associated}
\end{align}
with $d=\partial_t$ or $\delta$ by introducing one-coderivations
$\bd{\bar{\xi}_d}$ derived from $\bar{\xi}d$\,.
We can show that they satisfy the characteristic identities of the associated fields,
\begin{subequations}\label{associated id}
 \begin{align}
&\ dG_{\bar{\eta}}(V(t))\ =\
\pi_1^{(0,0)}\bd{L}^{\bar{\eta}}(e^{\wedge G_{\bar{\eta}}(V(t))}
\wedge B_d(V(t)))\,,\\
&\ 
D_{\bar{\eta}}(t)
\Big(
\partial_t B_\delta(V(t))-\delta  B_t(V(t))
\nonumber\\
&\hspace{3cm}
- \pi_1^{(0,0)}\bd{L}^{\bar{\eta}}(e^{\wedge G_{\bar{\eta}}(V(t))}\wedge
B_t(V(t))\wedge B_\delta(V(t)))
\Big)
\ =\ 0\,.
 \end{align}
\end{subequations}
The nilpotent linear operator
$D_{\bar{\eta}}(t)$ was introduced as
\begin{equation}
D_{\bar{\eta}}(t)\varphi\ =\ \pi_1^{(0,0)}\bd{L}^{\bar{\eta}}\Big(
e^{\wedge G_{\bar{\eta}}(V(t))}\wedge \varphi\Big)\,,
\end{equation}
for a general string field $\varphi\in\mathcal{H}_{\bar{m}}$\,.

The gauge transformation 
\begin{equation}
 \pi_1\delta(e^{\wedge\Phi})\ =\ \pi_1\bd{L}(e^{\wedge\Phi}\wedge\Lambda)
\end{equation}
generated by the parameter $\Lambda=\Lambda_{\NSNS}+\Lambda_{\RNS}
+\Lambda_{\NSR}+\Lambda_{\RR}$ is also mapped to that in the half-WZW-like formulation
with the gauge parameters
\begin{equation}
 \Lambda\ =\
  -\pi_1^{(0,0)}\hat{\bd{g}}(e^{\wedge\Phi_{\NSNS}}\wedge\bar{\xi}\Lambda_{\NSNS})\,,\quad
\lambda\ =\ \Lambda_{\RNS}\,,\quad \bar{\lambda}\ =\ \Lambda_{\NSR}\,,\quad
\rho\ =\ \Lambda_{\RR}\,,
\end{equation}
as
\begin{subequations} \label{WZW-like gauge tf 1}
\begin{align}
 B_\delta(V)\ =&\ \pi_1^{(0,0)}\tilde{\bd{L}}\Big(
e^{\wedge G_{\bar{\eta}}+\Psi+\bar{\Psi}\Sigma}\wedge
(\Lambda-\bar{\xi}\lambda-\bar{\xi}\bar{\lambda}-\bar{\xi}\rho)\Big)\,,\\
\delta\Psi\ =&\ 
- \pi_1^{(1,0)}\tilde{\bd{L}}\Big(
e^{\wedge G_{\bar{\eta}}+\Psi+\bar{\Psi}\Sigma}\wedge
(\bar{\eta}\Lambda-\lambda-\bar{\lambda}-\rho)\Big)\,,\\
\delta\bar{\Psi}\ =&\ 
- \pi_1^{(0,1)}\tilde{\bd{L}}\Big(
e^{\wedge G_{\bar{\eta}}+\Psi+\bar{\Psi}\Sigma}\wedge
(\bar{\eta}\Lambda-\lambda-\bar{\lambda}-\rho)\Big)\,,\\
\delta\Sigma\ =&\
- \pi_1^{(1,1)}\tilde{\bd{L}}\Big(
e^{\wedge G_{\bar{\eta}}+\Psi+\bar{\Psi}\Sigma}\wedge
(\bar{\eta}\Lambda-\lambda-\bar{\lambda}-\rho)\Big)\,,
\end{align}
\end{subequations}
where
\begin{equation}
\pi_1\tilde{\bd{L}}\ =\ \pi_1\hat{\bd{g}}\bd{L}\hat{\bd{g}}^{-1}\
=\ \pi_1\bd{Q} + \mathcal{G}\tilde{\bd{b}}\,.
\end{equation}
There is also an extra gauge invariance in the half-WZW-like formulation
under the transformation 
\begin{equation}
B_\delta(V)=D_{\bar{\eta}}\Omega\,,
\label{WZW-like gauge tf 2} 
\end{equation}
because the identification in 
Eq.~(\ref{identification NSNS}) is not one-to-one but
\begin{equation}
 \delta G_{\bar{\eta}}(V)\ =\ D_{\bar{\eta}}B_\delta(V)\ =\ 
D_{\bar{\eta}}D_{\bar{\eta}}\Omega\ =\ 0\,.
\end{equation}
The identities in Eq.~(\ref{associated id}) are enough to guarantee
that the action in Eq.~(\ref{WZW-like action}) is invariant under the gauge 
transformations in Eqs.~(\ref{WZW-like gauge tf 1}) and (\ref{WZW-like gauge tf 2})
independently with the gauge invariance in the homotopy algebraic formulation.

\section{Summary and discussion}\label{summary}

Extending the procedure for constructing the heterotic string field theory,
we have constructed the type II superstring field theory with a cyclic $L_\infty$
structure based on the homotopy algebraic formulation. 
In addition to the closed string constraints, we impose extra constraints on 
the string fields in the R-NS, NS-R and R-R sectors. 
These constraints restrict the dependence of these string fields on 
the bosonic ghost zero modes, and also make the field-anti-field decomposition 
in the BV quantization obvious.
Although the kinetic term of the R-R string field is non-local, it provides the same
propagator in the Siegel gauge as that naturally obtained in the first-quantized
formulation. 
Repeating the procedure
used in the construction of the heterotic string products, we have constructed
the string products for the type II superstring with the cyclic $L_\infty$
structure acting across all the NS-NS, R-NS, NS-R and R-R sectors.
We can map the action and the gauge transformation 
to those in the half-WZW-like 
formulation defined using the medium Hilbert space 
although not in the completely WZW-like formulation in the large Hilbert space.

A remaining interesting task is to construct a completely WZW-like action in the 
large Hilbert space. 
In the language introduced in Ref.~\cite{Matsunaga:2014wpa}, 
the similarity transformation generated by the cohomomorphism in Eq.~(\ref{cohomomorphism}) 
maps the small Hilbert space
$L_\infty$ triplet $(\bd{\eta},\bar{\bd{\eta}};\bd{L}^{\textrm{NS,NS}})$ 
to the asymmetric (heterotic) one 
$(\bd{\eta},\bd{L}^{\bar{\eta}};\bd{Q})$\,,
while the symmetric triplet $(\bd{L}^\eta,\bd{L}^{\bar{\eta}};\bd{Q})$ is 
necessary to realize the complete WZW-like formulation.
In order to realize it and couple the NS-NS action 
in Ref.~\cite{Matsunaga:2016zsu} to the string fields in the other sectors,
it seems to be necessary to find a construction which is an extension of 
the symmetric construction proposed in Ref.~\cite{Erler:2014eba}. 
Our method used in this paper, which was developed in Ref.~\cite{Kunitomo:2019glq} 
for constructing the heterotic string field theory, 
cannot be extended that way so some completely different 
approach seems to be needed.

Finally, it should be emphasized that the type II superstring field theory 
has the possibility of providing a solid basis to the AdS/CFT
correspondence which is still mysterious and must be proved.
We hope that the gauge-invariant action we have constructed will help us
explore such an interesting possibility.

\bigskip
\noindent
{\bf \large Acknowledgments}

The work of H.K. was supported in part by JSPS Grant-in-Aid for Scientific 
Research (C) number JP18K03645.

\appendix

\setcounter{equation}{0}
\section{Expanding with respect to ghost zero modes}
\label{ghost zero modes}

In this appendix we expand the string field with respect to 
the ghost zero modes.
After summarizing the Fock representation,
an explicit expression is given for each sector.

\subsection{Fock representation of ghost zero modes}
\label{Fock rep}

\subsubsection{fermionic ghost}

%
The fermionic ghost zero modes $(b_0,c_0)$ satisfy the anti-commutation
relation
\begin{equation}
 \{b_0\,,c_0\}\ =\ 1\,, 
\end{equation}
and the hermite and BPZ conjugate relations
\begin{alignat}{3}
 (b_0)^\dag\ =&\ b_0\,,\qquad & (c_0)^\dag\ =&\ c_0\,,\\
\textrm{bpz}(b_0)\ =&\ b_0\,, \qquad& 
\textrm{bpz}(c_0)\ =&\ -c_0\,.
\end{alignat}
Its Fock representation is two-dimensional space spanned by two states
\begin{equation}
 \{|\downarrow\,\rangle\,,\, |\uparrow\,\rangle\}\,,
\end{equation}
where
\begin{equation}
 b_0|\downarrow\,\rangle\ =\ 0\,,\qquad |\uparrow\,\rangle\ =\ c_0|\downarrow\,\rangle\,.
\end{equation}
Its dual space is spanned by their BPZ conjugates
\begin{equation}
 \{\langle\,\downarrow|\,,\, \langle\,\uparrow|\}\,,
\end{equation}
with
\begin{equation}
 \langle\,\downarrow|b_0\ =\ 0\,,\qquad \langle\,\uparrow|\ =\ \langle\,\downarrow|c_0\,.
\end{equation}
The inner-product matrix between two spaces is off-diagonal:
\begin{equation}
 \langle\,\uparrow|\downarrow\,\rangle\ =\ \langle\,\downarrow|c_0|\downarrow\,\rangle\ =\
\langle\,\downarrow|\uparrow\,\rangle\ =\ 1\,,\qquad
\langle\,\uparrow|\uparrow\,\rangle\ =\ \langle\,\downarrow|\downarrow\,\rangle\ =\ 0\,.
\end{equation}
Since there are two pairs of fermionic ghost zero modes, $(b_0,c_0)$ and 
$(\bar{b}_0,\bar{c}_0)$\,,
in the closed string theory, their Fock representation is four-dimensional
\begin{equation}
 \{|\downarrow\downarrow\,\rangle\,,\,
|\uparrow\downarrow\,\rangle\,,\,
|\downarrow\uparrow\,\rangle\,,\,
|\uparrow\uparrow\,\rangle\}\,,
\end{equation}
where $|\downarrow\downarrow\,\rangle\propto|\downarrow\,\rangle\otimes|\downarrow\,\rangle$\,.
One of the closed string constraints $b_0^-|\downarrow\downarrow\,\rangle=0$
restricts it to the two-dimensional space,
\begin{equation}
 \{|\downarrow\downarrow\,\rangle\,,\, c_0^+|\downarrow\downarrow\,\rangle\}\,.
\end{equation}
We normalize the states so that
\begin{equation}
 \langle\,\downarrow\downarrow|c_0^+c_0^-|\downarrow\downarrow\,\rangle\ =\ 1\,.
\end{equation}

\subsubsection{Bosonic ghost}

There are also bosonic ghosts $(\beta_n,\gamma_n)$ in the R sector.
They satisfy the commutation relation
\begin{equation}
[\gamma_n\,,\,\beta_m]\ =\ \delta_{n+m,0}\,, 
\end{equation}
and the hermite and BPZ conjugate relations\footnote{
Fractional phases in Eq.~(\ref{bpz beta-gamma}) become
a simple sign factor on the Fock states restricted by the GSO projection.}
\begin{alignat}{3} 
(\gamma_n)^\dag\ =&\ \gamma_{-n}\,,\qquad& (\beta_n)^\dag\ 
=&\ -\beta_{-n}\,,\\ 
\textrm{bpz}(\gamma_n)\ =&\ e^{-i\pi(n+\frac{1}{2})}\gamma_{-n}\,,\qquad&
\textrm{bpz}(\beta_n)\ =&\ e^{-i\pi(n-\frac{3}{2})}\beta_{-n}\,.
\label{bpz beta-gamma}
\end{alignat}
In general it is known that they 
have infinitely many Fock representations defined on the ground states
with the picture number $p$ as \cite{Friedan:1985ge} 

\begin{alignat}{3}
\beta_n|p\rangle\ =&\ 0\,,\qquad &  &\textrm{for}\ n> -p-\frac{3}{2}\,,\\
 \gamma_n|p\rangle\ =&\ 0\,,\qquad & &\textrm{for}\ n\ge p+\frac{3}{2}\,.
\end{alignat}

In string field theory
we use two natural representation with $p=-1/2$ and $-3/2$\,,
whose non-zero mode parts are common
Fock space obtained by acting $(\beta_{-n},\gamma_{-n})$ with
$n>0$ on the ground state $|0\rrangle$ satisfying
\begin{equation}
 \beta_n|0\rrangle\ =\ \gamma_n|0\rrangle\ =\ 0\,,\qquad n>0\,.
\end{equation}
Two representations with $p=-1/2$ and $-3/2$ are the direct product of 
this common non-zero mode part and the zero mode parts obtained by acting
$\gamma_0$ and $\beta_0$ on the ground states $|0\rangle$ and 
$|\tilde{0}\rangle$  defined by
\begin{equation}
\beta_0|0\rangle\ =\ 0\,,\qquad\qquad
\gamma_0|\tilde{0}\rangle\ =\ 0\,,
\end{equation}
respectively.
The representations of zero modes,
\begin{equation}
\{|0\rangle,\gamma_0|0\rangle,(\gamma_0)^2|0\rangle,\cdots\}
\quad \textrm{and}\quad
\{|\tilde{0}\rangle,\beta_0|\tilde{0}\rangle,(\beta_0)^2|\tilde{0}\rangle,\cdots\}\,,
\end{equation}
are infinite-dimensional and 
dual to each other with respect to the BPZ inner product induced by 
\begin{equation}
 \langle\tilde{0}|0\rangle\ =\ \langle 0|\tilde{0}\rangle\ =\ 1\,.
\label{zero mode dual}
\end{equation}
In order to intertwine two representations,
we can introduce the \textit{delta functions}
$\delta(\gamma_0)$ and $\delta(\beta_0)$ as 
the Grassmann odd operators
which satisfy
\begin{align}
 &\gamma_0\delta(\gamma_0)\ =\ \delta(\gamma_0)\gamma_0\ =\ 0\,,\qquad
 \beta_0\delta(\beta_0)\ =\ \delta(\beta_0)\beta_0\ =\ 0\,,\\
 &\delta(\beta_0)\delta(\gamma_0) \delta(\beta_0)\ =\  \delta(\beta_0)\,,\qquad
\delta(\gamma_0) \delta(\beta_0)\delta(\gamma_0)\ =\ \delta(\gamma_0)\,,
\end{align}
and are (graded) commutative with the operators other than $(\beta_0,\gamma_0)$\,.
Then, two ground states $|0\rangle$ and $|\tilde{0}\rangle$ can be related as
\begin{equation}
 |0\rangle\ =\ \delta(\beta_0)|\tilde{0}\rangle\,,\qquad
 |\tilde{0}\rangle\ =\ \delta(\gamma_0)|0\rangle\,.
\end{equation}
The inner products in Eq.~(\ref{zero mode dual}) provide
\begin{equation}
 \langle0|\delta(\gamma_0)|0\rangle\ =\ 
\langle\tilde{0}|\delta(\beta_0)|\tilde{0}\rangle\ =\ 1\,.
\end{equation}
These apparently strange operators can be defined by 
\begin{equation}
\delta(\gamma_0)\ =\ |\tilde{0}\rangle\langle\tilde{0}|\,,\qquad
\delta(\beta_0)\ =\ |0\rangle\langle0|\,,
\end{equation}
if necessary, and are closely related to a geometric object, 
the integral form on the super-moduli space of the super Riemann manifold
\cite{Belopolsky:1997bg,Belopolsky:1997jz}.\footnote{ The authors would like
to thank Pietro Antonio Grassi, from whom they learned much about the integral 
form \cite{Cremonini:2019aao,Catenacci:2019ksa,Catenacci:2018xsv}.}

\subsection{Zero-mode expansion of string fields}

\subsubsection{NS-NS sector}\label{NS NS}

In the NS-NS sector, only the fermionic ghosts have zero modes $(b_0,c_0)$
and $(\bar{b}_0,\bar{c}_0)$\,. 
The NS-NS string field $\Phi_{\NSNS}$ constrained by
Eq.~(\ref{restrict closed}) can be expanded 
with respect to these ghost zero modes as
\begin{equation}
 \Phi_{\NSNS}\ =\ \phi_{\NSNS} - c_0^+\psi_{\NSNS}\,.
\end{equation}
From its Fock representation,
we can separate the ghost zero-mode dependence as
\begin{equation}
 |\phi_{\NSNS}\rangle\ =\ |\downarrow\downarrow\,\rangle\otimes|\phi_{\NSNS}\rrangle\,,
\qquad
|\psi_{\NSNS}\rangle\ =\ |\downarrow\downarrow\,\rangle\otimes|\psi_{\NSNS}\rrangle\,,
\end{equation}
where the state denoted as $|\phi_{\NSNS}\rrangle$ or $|\psi_{\NSNS}\rrangle$
represents the non-zero mode part of the Fock representation of the string field.
This expansion holds independent of
whether the ghost number of $\Phi_{\NSNS}$ is restricted or not.

\subsubsection{R-NS sector}\label{R NS}

In the R-NS sector there are also the bosonic ghost zero modes $(\beta_0,\gamma_0)$
in the left-moving sector. 
The R-NS string field $\Phi_{\RNS}$ restricted by the constraints
in Eqs.~(\ref{restrict closed}) and (\ref{restrict RNS}) can be expanded as
\begin{equation}
\Phi_{\RNS}=\phi_{\RNS}-\frac{1}{2}(\gamma_0+2c_0^+G)\psi_{\RNS}\,,
\end{equation}
in which the ghost zero-mode dependence can be separated as
\begin{equation}
 |\phi_{\RNS}\rangle\ =\ 
|\downarrow\downarrow\,\rangle\otimes|0\rangle\otimes|\phi_{\RNS}\rrangle\,,\qquad
|\psi_{\RNS}\rangle\ =\ 
|\downarrow\downarrow\,\rangle\otimes|0\rangle\otimes|\psi_{\RNS}\rrangle\,.
\end{equation}
The states denoted as $|\phi_{\RNS}\rrangle$ and $|\psi_{\RNS}\rrangle$ are
the string field after separating the ghost-zero modes. 
The zero modes can be integrated out by using the inner product
\begin{equation}
 \langle 0|\otimes\langle\downarrow\downarrow|c_0^+c_0^-\delta(\gamma_0)
|\downarrow\downarrow\,\rangle\otimes|0\rangle\ =\ 1\,.
\end{equation}

\subsubsection{NS-R sector}\label{NS R}

In the NS-R sector the bosonic ghost zero modes
$(\bar{\beta}_0,\bar{\gamma}_0)$ are in the right-moving sector.
The NS-R string field $\Phi_{\NSR}$
restricted by the constraints 
in Eqs.~(\ref{restrict closed}) and (\ref{restrict RNS})
can be expanded as
\begin{equation}
 \Phi_{\NSR}\ =\ \phi_{\NSR} - \frac{1}{2}(\bar{\gamma}_0+2c_0^+\bar{G})\psi_{\NSR}\,,
\end{equation}
and we can separate the zero-mode part as
\begin{equation}
 |\phi_{\NSR}\rangle\ =\
  |\downarrow\downarrow\,\rangle\otimes|0\rangle\otimes|\phi_{\NSR}\rrangle\,,\qquad
 |\psi_{\NSR}\rangle\ =\
  |\downarrow\downarrow\,\rangle\otimes|0\rangle\otimes|\psi_{\NSR}\rrangle\,.
\end{equation}
%
The ghost zero modes can be integrated out using
\begin{equation}
 \langle0|\otimes\langle\,\downarrow\downarrow|c_0^+c_0^-\delta(\bar{\gamma}_0)
|\downarrow\downarrow\,\rangle\otimes|0\rangle\ =\ 1\,.
\end{equation}

\subsubsection{R-R sector}\label{R R}

In the R-R sector, there are the bosonic ghost zero modes 
$(\beta_0,\gamma_0)$ and $(\bar{\beta}_0,\bar{\gamma}_0)$ 
in both the left-moving and right-moving sectors.
The R-R string field restricted by the constraints
in Eqs.~(\ref{restrict closed}) and (\ref{constraint RR})
can be expanded as
\begin{equation}
 \Phi_{\RR}\ =\ \phi_{\RR} - \frac{1}{2}
(\gamma_0\bar{G}-\bar{\gamma}_0G+2c_0^+G\bar{G})\psi_{\RR}\,,
\end{equation}
in which we can further separate the ghost zero modes as
\begin{equation}
 |\phi_{\RR}\rangle\ =\ |\downarrow\downarrow\,\rangle\otimes
|0\rangle\otimes|0\rangle\otimes|\phi_{\RR}\rrangle\,,\qquad
 |\psi_{\RR}\rangle\ =\ |\downarrow\downarrow\,\rangle\otimes
|0\rangle\otimes|0\rangle\otimes|\psi_{\RR}\rrangle\,,
\end{equation}
with the string fields $|\phi_{\RR}\rrangle$ and $|\psi_{\RR}\rrangle$ 
constructed only on the non-zero-mode Fock space.
The zero modes can be integrated out by using the non-trivial inner product
\begin{equation}
 \langle0|\otimes\langle0|\otimes\langle\,\downarrow\downarrow|c_0^+c_0^-\delta(\gamma_0)
\delta(\bar{\gamma}_0)|\downarrow\downarrow\,\rangle\otimes|0\rangle\otimes|0\rangle\ =\ 1\,.
\end{equation}

In order to construct the Sen-type action, we have to introduce an extra
string field with picture number $(-3/2,-3/2)$\,,
whose zero-mode ground state 
$ |\downarrow\downarrow \rangle\otimes|\tilde{0}\rangle\otimes|\tilde{0}\rangle $ 
is related to
$|\downarrow\downarrow \rangle \otimes|0\rangle\otimes|0\rangle$
through the relations
\begin{align}
 |\downarrow\downarrow\,\rangle\otimes|0\rangle\otimes|0\rangle\ =&\ \delta(\bar{\beta}_0)
\delta(\beta_0)|\downarrow\downarrow\,\rangle\otimes|\tilde{0}\rangle\otimes
|\tilde{0}\rangle\,\\
 |\downarrow\downarrow\,\rangle\otimes|\tilde{0}\rangle\otimes|\tilde{0}\rangle\
=&\ \delta(\gamma_0)
\delta(\bar{\gamma}_0)|\downarrow\downarrow\,\rangle\otimes|0\rangle\otimes
|0\rangle\,.
\end{align}
Their BPZ conjugates are
\begin{align}
\langle0|\otimes\langle0|\otimes\langle\,\downarrow\downarrow|\ =\ 
\langle\tilde{0}|\otimes\langle\tilde{0}|\otimes
\langle\,\downarrow\downarrow|\delta(\bar{\beta}_0)\delta(\beta_0)\,,\\
\langle\tilde{0}|\otimes\langle\tilde{0}|\otimes\langle\,\downarrow\downarrow|\ =\ 
\langle0|\otimes\langle0|\otimes
\langle\,\downarrow\downarrow|\delta(\gamma_0)\delta(\bar{\gamma}_0)\,.
\end{align}
The zero-mode integration can be performed by using the inner product
\begin{equation}
 \langle\tilde{0}|\otimes\langle\tilde{0}|\otimes
\langle\,\downarrow\downarrow|c_0^+c_0^-\delta(\bar{\beta}_0)
\delta(\beta_0)|\downarrow\downarrow\,\rangle\otimes|\tilde{0}\rangle
\otimes|\tilde{0}\rangle\ =\ 1\,.
\end{equation}
The extra string field $\tilde{\Phi}_{\RR}$ restricted by the constraint
in Eq.~(\ref{constraint tilde RR}) can be expanded with respect to the ghost zero modes as
\begin{equation}
 \tilde{\Phi}_{\RR}\ =\ \tilde{\phi}_{\RR} - c_0^+\tilde{\psi}_{\RR}\,,
\end{equation}
from which we can separate the ghost zero modes as
\begin{subequations} 
\begin{align}
 \tilde{\phi}_{\RR}\ =&\ 
|\downarrow\downarrow\,\rangle\otimes|\tilde{0}\rangle\otimes
|\tilde{0}\rangle\otimes|\tilde{\phi}_{\RR}\rrangle\,,\\
 \tilde{\psi}_{\RR}\ =&\ 
|\downarrow\downarrow\,,\rangle\otimes|\tilde{0}\rangle\otimes
|\tilde{0}\rangle\otimes|\tilde{\psi}_{\RR}\rrangle\,.
\end{align}
\end{subequations}
It can be shown that $X\bar{X}\tilde{\Phi}_{\RR}$ obtained by changing 
the picture actually has the form of Eq.~(\ref{restricted RR}):
\begin{equation}
 X\bar{X}\tilde{\Phi}_{\RR}\ =\ G\bar{G}\tilde{\tilde{\phi}}_{\RR}
- \frac{1}{2}(\gamma_0\bar{G} - \bar{\gamma}_0G + 2c_0^+G\bar{G})\tilde{\tilde{\psi}}_{\RR}\,,
\end{equation}
where 
\begin{subequations} 
 \begin{align}
\tilde{\tilde{\phi}}_{\RR}\ =&\
|\downarrow\downarrow\,\rangle\otimes|0\rangle\otimes
|0\rangle\otimes|\tilde{\phi}_{\RR}\rrangle\,,\\
\tilde{\tilde{\psi}}_{\RR}\ =&\ 
|\downarrow\downarrow\,\rangle\otimes
|0\rangle\otimes|0\rangle\otimes|\tilde{\psi}_{\RR}\rrangle\,.
\end{align}
\end{subequations}

\end{document}